\documentclass[aps,preprint,amsmath,amssymb]{revtex4}
\usepackage{graphicx}
\begin{document}

\title{Analysis of the anomalous quartic $WWWW$ couplings at the LHeC and the FCC-he}

\author{E. Gurkanli}
\email[]{egurkanli@sinop.edu.tr}\affiliation{Department of Physics, Sinop University, Turkey}

\author{V. Ari}
\email[]{vari@science.ankara.edu.tr} \affiliation{Department of Physics, Ankara University, Turkey}

\author{A. A. Billur}
\email[]{abillur@cumhuriyet.edu.tr} \affiliation{Deparment of Physics, Sivas Cumhuriyet University, Turkey}

\author{M. K\"{o}ksal}
\email[]{mkoksal@cumhuriyet.edu.tr} \affiliation{Department of Optical Engineering, Sivas Cumhuriyet University, Turkey}

\begin{abstract}
The quartic gauge boson couplings that identify the strengths of the gauge boson self-interactions are exactly determined by the non-Abelian
gauge nature of the Standard Model. The examination of these couplings at $ep$ collisions with high center-of-mass energy and high integrated luminosity provides an important opportunity to test the validity of the Standard Model and the existence of new physics beyond the Standard Model. The quartic gauge boson couplings can contribute directly to multi-boson production at colliders. Therefore, we examine the potential of the process $ep \rightarrow \nu_{e}W^{+}W^{-} j$ at the Large Hadron Electron Collider and the Future Circular Collider-hadron electron to study non-standard $WWWW$ couplings in a model independent way by means of the effective Lagrangian approach. We present an investigation on measuring $W^{+}W^{-}$ production in pure leptonic and semileptonic decay channels.
In addition, we calculate the sensitivity limits at $95\%$ Confidence Level on the anomalous $\frac{f_{M0}}{\Lambda^{4}}$, $\frac{f_{M1}}{\Lambda^{4}}$, $\frac{f_{M7}}{\Lambda^{4}}$, $\frac{f_{S0}}{\Lambda^{4}}$, $\frac{f_{S1}}{\Lambda^{4}}$, $\frac{f_{T0}}{\Lambda^{4}}$, $\frac{f_{T1}}{\Lambda^{4}}$ and $\frac{f_{T2}}{\Lambda^{4}}$ couplings obtained by dimension-8 operators through the process $ep \rightarrow \nu_{e}W^{+}W^{-} j$ for the Large Hadron Electron Collider and the Future Circular Hadron Electron Collider's different center-of-mass energies and integrated luminosities. Our results show that with the process $ep \rightarrow \nu_{e}W^{+}W^{-} j$ at the Large Hadron Electron Collider and the Future Circular Collider-hadron electron the sensitivity estimated on the anomalous $WWWW$ couplings can be importantly strengthened.
\end{abstract}

\maketitle

\section{Introduction}

Since the Large Hadron Collider (LHC) began to receive data, there is no significant deviation from the Standard Model (SM). Instead, with the ultimate discovery of the approximately 125 GeV Higgs boson at the LHC in 2012, the SM has achieved an important success \cite{higgs1,higgs2}. Even so, new physics beyond the SM is needed to clarify the deficiencies of the SM such as neutrino oscillations, the strong CP problem, and matter-antimatter asymmetry. The examination of the gauge boson self-interactions is important for the precise testing of SM and new physics research beyond the SM. Thus, it is possible to carry out different measurements in processes involving multi-boson production in colliders with high center-of-mass energy and high luminosity. However, contributions to the quartic gauge boson couplings in the SM that can arise from new physics can be parametrized in a model independent way through the effective Lagrangian approach. These operators that define the anomalous quartic gauge boson couplings are based on the non-linear or linear representation realization of the gauge symmetry \cite{bel1,bel2}.

It is assumed that there is no Higgs boson at low energy spectrum in the non-linear representation. Besides, with the discovery of the Higgs particle, a linear representation of gauge symmetry that is broken by the conventional Higgs mechanism is probable. Therefore, in linear representation, the lowest order operators that define the possible deviations of the quartic gauge boson couplings from the SM are dimension-8 \cite{baa}:

\begin{eqnarray}
\mathcal{L}_{eff}=\sum_{i=1}^{2}\frac{f_{Si}}{\Lambda^{4}}\uppercase{o}_{Si}+\sum_{j=0,1,2,5,6,7,8,9}\frac{f_{Tj}}{\Lambda^{4}}\uppercase{o}_{Tj}+\sum_{k=0}^{7}\frac{f_{Mk}}{\Lambda^{4}}\uppercase{o}_{Mk},
\end{eqnarray}
where the $\uppercase{o}_{n}$ operators possess $f_{n}$ couplings and $\Lambda$ is a characteristic scale. There are 17 different operators that define the anomalous quartic gauge boson couplings. The indices \textit{S}, \textit{T}, and \textit{M} of the operators depict three different classes: operators including only $D_{\mu}\Phi$, operators containing $D_{\mu}\Phi$ and field strength and operators exhibiting four field strength tensors.

The first class has two independent operators as follows

\begin{eqnarray}
\uppercase{o}_{S0}&=&[(D_{\mu}\Phi)^{\dagger} D_{\nu}\Phi] \times [(D^{\mu}\Phi)^{\dagger} D^{\nu}\Phi],\\
\uppercase{o}_{S1}&=&[(D_{\mu}\Phi)^{\dagger}D^{\mu}\Phi]\times [(D_{\nu}\Phi)^{\dagger}D^{\nu}\Phi],
\end{eqnarray}
where $\Phi$ is Higgs doublet field.

The eight operators of second class are given by

\begin{eqnarray}
\uppercase{o}_{M0}&=&Tr[W_{\mu\nu}W^{\mu\nu}]\times[(D_{\beta}\Phi)^{\dagger}D^{\beta}\Phi],\\
\uppercase{o}_{M1}&=&Tr[W_{\mu\nu}W^{\nu\beta}]\times[(D_{\beta}\Phi)^{\dagger}D^{\mu}\Phi],\\
\uppercase{o}_{M2}&=&Tr[B_{\mu\nu}B^{\mu\nu}]\times[(D_{\beta}\Phi)^{\dagger}D^{\beta}\Phi],\\
\uppercase{o}_{M3}&=&Tr[B_{\mu\nu}B^{\nu\beta}]\times[(D_{\beta}\Phi)^{\dagger}D^{\mu}\Phi],\\
\uppercase{o}_{M4}&=&[(D_{\mu}\Phi)^{\dagger}W_{\beta\nu} D^{\mu}\Phi]\times B^{\beta\nu},\\
\uppercase{o}_{M5}&=&[(D_{\mu}\Phi)^{\dagger}W_{\beta\nu} D^{\nu}\Phi]\times B^{\beta\mu},\\
\uppercase{o}_{M6}&=&[(D_{\mu}\Phi)^{\dagger}W_{\beta\nu}W^{\beta\nu} D^{\mu}\Phi],\\
\uppercase{o}_{M7}&=&[(D_{\mu}\Phi)^{\dagger}W_{\beta\nu}W^{\beta\mu} D^{\nu}\Phi].
\end{eqnarray}

Here, $W_{\mu\nu}$ and $B_{\mu}$ that are the electroweak field strength tensors are given as follows
\begin{eqnarray}
W_{\mu\nu}&=&\frac{i}{2} g \tau^i ( \partial_{\mu} W_{\nu}^i-\partial_{\nu}W_{\mu}^i+g \epsilon_{ijk} W_{\mu}^j W_{\nu}^k),\nonumber\\
B_{\mu\nu}&=&\frac{i}{2}g'(\partial_{\mu}B_{\nu}-\partial_{\nu}B_{\mu}),
\end{eqnarray}
where $e$ stands for the unit of electric charge and $\theta_W$ is the Weinberg angle, $\tau^i (i=1,2,3)$ represents the $SU(2)$ generators, $g$ and $g'$ are coupling constant of $U(1)$ and $SU(2)$, respectively.

The final class is expressed as follows

\begin{eqnarray}
\uppercase{o}_{T0}&=& Tr[W_{\mu\nu}W^{\mu\nu}]\times [W_{\alpha\beta}W^{\alpha\beta}],\\
\uppercase{o}_{T1}&=&Tr[W_{\alpha\nu}W^{\mu\beta}]\times [W_{\mu\beta}W^{\alpha\nu}],\\
\uppercase{o}_{T2}&=&Tr[W_{\alpha\mu}W^{\mu\beta}]\times [W_{\beta\nu}W^{\nu\alpha}],\\
\uppercase{o}_{T5}&=&Tr[W_{\mu\nu}W^{\mu\nu}]\times B_{\alpha\beta}B^{\alpha\beta},\\
\uppercase{o}_{T6}&=&Tr[W_{\alpha\nu}W^{\mu\beta}]\times B_{\beta\mu}B^{\alpha\nu},\\
\uppercase{o}_{T7}&=&Tr[W_{\alpha\mu}W^{\mu\beta}]\times B_{\beta\nu}B^{\nu\alpha},\\
\uppercase{o}_{T8}&=&[B_{\mu\nu}B^{\mu\nu}B_{\alpha\beta}B^{\alpha\beta}],\\
\uppercase{o}_{T9}&=&[B_{\alpha\nu}B^{\mu\beta}B_{\beta\nu}B^{\nu\alpha}].
\end{eqnarray}

Table I shows the changes in the quartic gauge boson couplings in the SM caused by the above operators.

Many theoretical and experimental studies have been carried out for processes involving the anomalous quartic gauge boson couplings via dimension-6 and dimension-8 operators \cite{x1,x2,x3,x4,x5,x6,x7,x8,x9,x91,x92,x10,x11,x12,x13,x14,x141,x15,x16,x17,x18,x19,x21,x22,x33,yy,zz,pp,sir1,sir2,sir3,sir4,ma1,ma2,ma3,sir5,11,12}. However, in this study, we consider $\mathcal{L}_{M0}$, $\mathcal{L}_{M1}$, $\mathcal{L}_{M7}$, $\mathcal{L}_{S0}$, $\mathcal{L}_{S1}$, $\mathcal{L}_{T0}$, $\mathcal{L}_{T1}$ and $\mathcal{L}_{T2}$ effective Lagrangians which contribute to the anomalous quartic $WWWW$ couplings. As can be seen in Table V, $\frac{f_{M0}}{\Lambda^{4}}$, $\frac{f_{M1}}{\Lambda^{4}}$, $\frac{f_{M7}}{\Lambda^{4}}$, $\frac{f_{S0}}{\Lambda^{4}}$, $\frac{f_{S1}}{\Lambda^{4}}$, $\frac{f_{T0}}{\Lambda^{4}}$, $\frac{f_{T1}}{\Lambda^{4}}$ and $\frac{f_{T2}}{\Lambda^{4}}$ couplings can be examined via $\mathcal{L}_{M0}$, $\mathcal{L}_{M1}$, $\mathcal{L}_{M7}$, $\mathcal{L}_{S0}$, $\mathcal{L}_{S1}$, $\mathcal{L}_{T0}$, $\mathcal{L}_{T1}$ and $\mathcal{L}_{T2}$ effective Lagrangians, respectively. In this study, we do not considered $\mathcal{L}_{M6}$ effective Lagrangian because it is not implemented in the UFO model files \cite{web}.

In the literature, the ATLAS Collaboration at the LHC has investigated $\frac{f_{S0}}{\Lambda^4}$ and $\frac{f_{S1}}{\Lambda^4}$ couplings for triboson production in two decay channels \cite{atlas}. Also, the CMS Collaboration sets the $\frac{f_{T0}}{\Lambda^{4}}$ , $\frac{f_{T1}}{\Lambda^{4}}$ and $\frac{f_{T2}}{\Lambda^{4}}$ couplings for the process $pp \rightarrow WWW$ with a center-of-mass energy of 13 TeV and a total integrated luminosity of 35.9 $fb^{-1}$ \cite{sir1}. In Ref. \cite{ebo}, the sensitivity limits on $\frac{f_{S0}}{\Lambda^4}$ and $\frac{f_{S1}}{\Lambda^4}$ couplings have been examined at $99\%$ Confidence Level for $ \sqrt{s}= 14$ TeV through the reaction $pp\rightarrow jje^{\pm}\mu^{\pm}\nu\nu$. As can be seen from the results of Snowmass, the sensitivity limits on $\frac{f_{T0}}{\Lambda^4}$ coupling with $95\%$ Confidence Level in pure leptonic channel through the triboson production have been studied \cite{deg}. In Ref. \cite{yu}, $\frac{f_{S0}}{\Lambda^4}, \frac{f_{S1}}{\Lambda^4}$ and $\frac{f_{T0}}{\Lambda^4}$ couplings have been investigated through the process $pp\rightarrow WWW$ at the LHC with $ \sqrt{s}= 14$ TeV and at the future hadron colliders with $ \sqrt{s}= 100$ TeV. All of the obtained limits on the anomalous $\frac{f_{S0}}{\Lambda^4}, \frac{f_{S1}}{\Lambda^4}$ and $\frac{f_{T0}}{\Lambda^4}$ couplings in these studies are given in Tables II-III.

Because of high background arising from QCD interactions, $pp$ colliders such as LHC may not be convenient for the precise measurements. In this regard, \textit{ep} colliders offer a cleaner environment than LHC. Therefore, new physics beyond the SM can be examined with $ep$ colliders with high center-of-mass energy and high luminosity. The Large Hadron electron Collider (LHeC) \cite{lhec,lhec1,lhec2,lhec3,lhec4,lhec5} and Future Circular Collider-hadron electron (FCC-he) \cite{fcc,fcc1,fcc2} are planned for the new generation $ep$ colliders in the near future projects. The future $ep$ colliders are considered to produce electron-proton collisions at center-of-mass energies from 1.30 TeV to 10 TeV.

This study presents a research on di-boson production through discovering the potential of measuring $W^{+}W^{-}$ final state with full leptonic decay and semileptonic decay channels at 1.30,1.98 TeV LHeC and 7.07,10 TeV FCC-he and examining the anomalous quartic $WWWW$ coupling. The values of $\sqrt{s}$ and $L_{int}$ used in the calculations are a little different from the ones in the FCC-CDR \cite{fcc1,fcc2}. On the other hand, the simulations have been done with the values given in  Ref. \cite{fcc}.

The remaining part of the study is planned as follows: in the following Section, the cross sections of the process $ep \rightarrow \nu_{e}W^{+}W^{-} j$ at the LHeC and the FCC-he are obtained. We focus on model independent sensitivity estimates on the anomalous quartic $WWWW$ couplings in Section III. Conclusions are discussed in Section 4.

\section{The cross section of the process $ep \rightarrow \nu_{e}W^{+}W^{-} j$}

In the production of signal process, the effective Lagrangians with the anomalous quartic couplings \cite{ebo,x141,x13} are implemented to FeynRules package \cite{rul} and embedded into MadGraph$5_{-}$aMC$@$NLO \cite{mad} as a Universal FeynRules Output \cite{ufo}. For parton distribution functions, the NN23LO1-PDF is adopted \cite{5-1,5-2,5-3,5-4}. Considering hadrons as a unified structure consisting of many sub-parts called partons, collision of two hadrons naturally involves collisions of many particles. This type of collision not only complicates the collision kinematics, but also makes it difficult to distinguish the desired signals in this case. The main reason for this phenomenon is that, unlike lepton collisions, hadron-hadron collisions involve many point 
particle collisions. However, lepton-hadron collisions mitigate these problems to some extent \cite{9-1}.
In order to investigate the possibilities of $ep$ colliders as an option to sensitivity estimates on the anomalous quartic $WWWW$ couplings, we concentrate $ep \rightarrow \nu_{e}W^{+}W^{-} j$ signal processes.

The squared amplitude for the process $ep \rightarrow \nu_{e}W^{+}W^{-} j$ with pure leptonic and semileptonic decay of $W$-bosons is given as follows:

\begin{eqnarray}
|M|^2=|M_{SM}|^{2}+2\Re(M_{SM}M^{*}_{dim8})+|M_{dim8}|^2
\end{eqnarray}
The first term is the contribution coming from the SM. The second term may contribute since the interference takes the contribution proportional to $\frac{f_{i}}{\Lambda^{4}}$. The last term includes the contribution of pure dim-8 due to the factor $\frac{f_{i}^{2}}{\Lambda^{8}}$ where $\Lambda$ is the high energy scale. The total cross section of the process $ep \rightarrow \nu_{e}W^{+}W^{-} j$ with the pure leptonic and semileptonic final states can be written as $\sigma_{tot}=\sigma_{SM}+\sigma_{int}+\sigma_{dim8}$ where the $\sigma_{SM}$ is the SM cross section with the leading order. On the other hand $\sigma_{int}$ is the interference between SM and the dim-8 term which is also a dim-8 contribution. Lastly, $\sigma_{dim8}$ represents the purely dim-$8^{2}$ contribution. We have shown the fit functions of the process for some anomalous quartic gauge couplings in the appendix.

In the analysis, we apply the following set of cuts in order to suppress the backgrounds and enhance the signal in the process $ep \rightarrow \nu_{e}W^{+}W^{-} j$ including the anomalous quartic $WWWW$ vertex. For pure leptonic decay channel, these cuts are given by

\begin{eqnarray}
p_{T_{j}}>20 GeV,p_{T_{\ell}}>10 GeV,
\end{eqnarray}

\begin{eqnarray}
 |\eta_{j}|< 5,|\eta_{\ell}|< 2.5,
\end{eqnarray}

\begin{eqnarray}
\Delta R(l,l)>0.4,\Delta R(j,l)>0.4
\end{eqnarray}

For semileptonic decay channel, applied cuts are

\begin{eqnarray}
p_{T_{j}}>20 GeV,p_{T_{\ell}}>10 GeV,
\end{eqnarray}

\begin{eqnarray}
 |\eta_{j}|< 5,|\eta_{\ell}|< 2.5,
\end{eqnarray}

\begin{eqnarray}
\Delta R(j,l)>0.4,\Delta R(j,j)>0.4,
\end{eqnarray}
where $\eta$ is the pseudorapidity, $p_T$ and $\Delta R$ are the transverse momentum and the separation of the final state particles, respectively. As can be seen in Table IV, with the effect of the cuts, the total cross sections including anomalous quartic gauge couplings decrease by about 50 percent while the decrease in the SM cross section is approximately 90 percent. On the other hand, for the semileptonic final state at $\sqrt{s}=1.98$ TeV in Table V, the decrease in SM cross section is approximately 70 percent but the decrease of the total cross sections including the anomalous quartic gauge couplings reaches approximately 50 percent.

\begin{table}
\caption{The anomalous quartic couplings modified by each dimension-8 effective Lagrangian are marked with X.\label{tab1}}
\begin{ruledtabular}
\begin{tabular}{ccccccccccccc}
&&$WWWW$ &$WWZZ$ & $ZZZZ$ & $WW\gamma Z$ & $WW\gamma \gamma$ & $ZZZ\gamma$ & $ZZ \gamma \gamma$ & $Z\gamma \gamma \gamma$ & $\gamma\gamma\gamma\gamma$ &\\
\hline
$\mathcal{L}_{S0},\mathcal{L}_{S1}$ &&$X$  &$X$ &$X$ & \\
\hline
$\mathcal{L}_{M0},\mathcal{L}_{M1},\mathcal{L}_{M6},\mathcal{L}_{M7}$ &&$X$ &$X$ &$X$ &$X$ &$X$ &$X$ &$X$  &\\
\hline
$\mathcal{L}_{M2},\mathcal{L}_{M3},\mathcal{L}_{M4},\mathcal{L}_{M5}$&&&$X$ &$X$ &$X$ &$X$ &$X$ &$X$ & \\
\hline
$\mathcal{L}_{T0},\mathcal{L}_{T1},\mathcal{L}_{T2}$&&$X$ &$X$ &$X$ &$X$ &$X$ &$X$ &$X$ &$X$ &$X$ & \\
\hline
$\mathcal{L}_{T5},\mathcal{L}_{T6},\mathcal{L}_{T7}$&&&$X$ &$X$ &$X$ &$X$ &$X$ &$X$ &$X$ &$X$ &\\
\hline
$\mathcal{L}_{T8},\mathcal{L}_{T9}$ &&&&$X$&& &$X$ &$X$ &$X$ &$X$ & \\
\hline
\end{tabular}
\end{ruledtabular}
\end{table}

\begin{table}
\caption{Limits on the anomalous $\frac{f_{S0}}{\Lambda^4}, \frac{f_{S1}}{\Lambda^4}$ and $\frac{f_{T0}}{\Lambda^4}$ parameters that determine the anomalous quartic $WWWW$ couplings in the literature.}
\begin{ruledtabular}
\begin{tabular}{cccccccccc}
$$&$L(fb^{-1})$&$\sqrt{s}$ (TeV)&$\frac{f_{S0}}{\Lambda^4}$(TeV$^{-4}$) &$\frac{f_{S1}}{\Lambda^4}$(TeV$^{-4}$) & $\frac{f_{T0}}{\Lambda^4}$(TeV$^{-4}$) &\\
\hline
\cite{atlas}&$20.3$ &8&$[-0.13; 0.18]\times 10^{4}$ &$[-0.21; 0.27]\times 10^{4}$ & $-$ &\\
\hline
\cite{ebo}&$100$ &14&$[-2.20; 2.40]\times 10^{1}$ &$[-2.50; 2.50]\times 10^{1}$ &$-$ &\\
\hline
\cite{yu}&$100$ &14&$[-1.80; 1.80]\times 10^{2}$ &$[-2.70; 2.80]\times 10^{2}$ &$[-0.58; 0.59]$ &\\
\hline
\cite{deg}&$300$ &14&$-$ &$-$ &$[-1.20; 1.20]$ &\\
\hline
\cite{deg}&$3000$ &14&$-$ &$-$ &$[-0.60; 0.60]$ &\\
\hline
\cite{deg}&$3000$ &33&$-$ &$-$ &$[-5.00; 5.00]\times 10^{-2}$ &\\
\hline
 \cite{deg}&$1000$ &100&$-$ &$-$ &$[-4.00; 4.00]\times 10^{-3}$ &\\
\hline
\cite{yu}&$3000$ &100&$[-2.93; 3.04]$ &$[-1.30; 1.16]$ &$[-3.69; 2.97]\times 10^{-3}$ &\\
\hline
 \cite{deg}&$3000$ &100&$-$ &$-$ &$[-2.00; 2.00]\times 10^{-3}$ &\\
\hline
\end{tabular}
\end{ruledtabular}
\end{table}

\begin{table}
\caption{Limits on the anomalous $\frac{f_{M0}}{\Lambda^4},\frac{f_{M1}}{\Lambda^4},\frac{f_{M7}}{\Lambda^4},\frac{f_{S0}}{\Lambda^4},\frac{f_{S1}}{\Lambda^4}, \frac{f_{T0}}{\Lambda^4}, \frac{f_{T1}}{\Lambda^4}$ and $\frac{f_{T2}}{\Lambda^4}$ parameters that determine the anomalous quartic $WWWW$ couplings in the literature.}
\begin{ruledtabular}
\begin{tabular}{cccccccccc}
$$&&$W\gamma$ \cite{19-1}&$Z\gamma$ \cite{19-2}& $WZ$, $WW$ \cite{19-3} &$W^{\pm}W^{\pm}W^{\pm}$ \cite{19-4} & 4-charged leptons+jj \cite{19-5} &\\
\hline
$\frac{f_{M0}}{\Lambda^4}$ &&$[-8.07; 7.99]$&$[-19.3; 20.2]$ &$[-2.70; 2.90]$ & $-$ &$-$&\\
\hline
$\frac{f_{M1}}{\Lambda^4}$& &$[-11.8; 12.1]$&$[-47.8; 46.9]$ &$[-4.10; 4.20]$ &$-$ &$-$&\\
\hline
$\frac{f_{M7}}{\Lambda^4}$& &$[-20.8; 20.2]$&$[-60.8; 62.6]$ &$[-5.40; 5.80]$ &$-$ &$-$&\\
\hline
$\frac{f_{S0}}{\Lambda^4}$& &$-$&$-$ &$[-5.70; 6.10]$ &$-$ &$-$&\\
\hline
$\frac{f_{S1}}{\Lambda^4}$& &$-$&$-$ &$[-16.0; 17.0]$ &$-$ &$-$&\\
\hline
$\frac{f_{T0}}{\Lambda^4}$& &$[-0.62; 0.64]$&$[-0.74; 0.69]$ &$[-0.25; 0.28]$ &$[-1.20; 1.20]$ &$[-0.24; 0.22]$&\\
\hline
$\frac{f_{T1}}{\Lambda^4}$& &$[-0.35; 0.39]$&$[-1.16; 1.15]$ &$[-0.12; 0.14]$ &$[-3.30; 3.30]$ &$[-0.31; 0.31]$&\\
\hline
$\frac{f_{T2}}{\Lambda^4}$& &$[-0.99; 1.18]$&$[-1.96; 1.85]$ &$[-0.35; 0.48]$ &$[-2.70; 2.60]$ &$[-0.63; 0.59]$&\\
\hline
\end{tabular}
\end{ruledtabular}
\end{table}

\begin{table}
\caption{Total cross sections and cut efficiencies for $\frac{f_{M0}}{\Lambda^{4}}$ , $\frac{f_{S0}}{\Lambda^{4}}$, $\frac{f_{T0}}{\Lambda^{4}}$ and the SM for leptonic final state at $\sqrt{s}=1.98$ TeV. Here, the values of anomalous quartic gauge couplings are taken as $1\times10^{-9}$ GeV$^{-4}$. }
\begin{center}
\begin{tabular}{|c|c|c|c|}
\hline
Couplings & No Cut (pb) & Cut(pb) & Efficiency ((No Cut)/Cut)\\
\hline
$\frac{f_{M0}}{\Lambda^{4}}$ & $0.0974$ & $0.0617$ & $0.63$  \\
\hline
$\frac{f_{S0}}{\Lambda^{4}}$ & $0.0036$ & $0.00168$ & $0.47$  \\
\hline
$\frac{f_{T0}}{\Lambda^{4}}$ & $4.118$ & $2.394$ & $0.58$\\
\hline
$SM$ & $0.0192$ & $0.0016$ & $0.08$\\
\hline
\end{tabular}
\end{center}
\end{table}

\begin{table}
\caption{Total cross sections and cut efficiencies for $\frac{f_{M0}}{\Lambda^{4}}$ , $\frac{f_{S0}}{\Lambda^{4}}$, $\frac{f_{T0}}{\Lambda^{4}}$ and the SM for semileptonic final state at $\sqrt{s}=1.98$ TeV. Here, the values of anomalous quartic gauge couplings are taken as $1\times10^{-9}$ GeV$^{-4}$.}
\begin{center}
\begin{tabular}{|c|c|c|c|}
\hline
Couplings & No Cut (pb) & Cut(pb) & Efficiency ((No Cut)/Cut)\\
\hline
$\frac{f_{M0}}{\Lambda^{4}}$ & $0.306$ & $0.143$ & $0.47$  \\
\hline
$\frac{f_{S0}}{\Lambda^{4}}$ & $0.0106$ & $0.00394$ & $0.37$  \\
\hline
$\frac{f_{T0}}{\Lambda^{4}}$ & $11.61$ & $5.733$ & $0.49$\\
\hline
$SM$ & $5.54\times10^{-5}$ & $1.57\times10^{-5}$ & $0.28$\\
\hline
\end{tabular}
\end{center}
\end{table}

At the LHeC and the FCC-he, $W^{+}W^{-}$ production can be generated via the process $ep \rightarrow \nu_{e}W^{+}W^{-} j$, the schematic diagram of this process is represented in Fig. 1. The total Feynman diagrams of the processes  $ep \rightarrow \nu_{e}W^{+} W^{-}j\rightarrow \nu_{e} \ell^{+}\nu_{\ell}\ell^{-}\bar\nu_{\ell}j$ and $ep \rightarrow \nu_{e}W^{+} W^{-}j\rightarrow \nu_{e} \ell^{+}\nu_{\ell}\ j j j$ are 248 where $l$ represents the light leptons ($e$,$\mu$) as the final-state leptons that are massless. For pure leptonic and semileptonic decay channels, as shown in Figs. 2-9, the total cross sections of the process $ep \rightarrow \nu_{e}W^{+}W^{-} j$ in terms of the anomalous $\frac{f_{M0}}{\Lambda^{4}}$, $\frac{f_{M1}}{\Lambda^{4}}$, $\frac{f_{M7}}{\Lambda^{4}}$, $\frac{f_{S0}}{\Lambda^{4}}$, $\frac{f_{S1}}{\Lambda^{4}}$, $\frac{f_{T0}}{\Lambda^{4}}$, $\frac{f_{T1}}{\Lambda^{4}}$ and $\frac{f_{T2}}{\Lambda^{4}}$ couplings for the LHeC and the FCC-he are displayed. The results demonstrate a clear dependence of the cross section of the process $ep \rightarrow \nu_{e}W^{+}W^{-} j$ with respect to the anomalous quartic couplings and the center-of-mass energies of the LHeC and the FCC-he. Here, we use a fixed factorization scale equal to $M_{Z}$ for all the numerical simulations.

The signal cross sections for a certain value of the anomalous couplings have been shown in Tables VI-IX. These tables show the possible deviations from the SM cross section for each anomalous coupling. In these tables, the value of the anomalous coupling values have been taken as  $1\times 10^{-8}$ GeV$^{-4}$. As can be seen in the tables, the largest deviation from the SM cross section for the anomalous couplings is $\frac{f_{T0}}{\Lambda^{4}}$. In this case, we expect that the obtained limits on $\frac{f_{T0}}{\Lambda^{4}}$  parameter should be more restrictive than the other parameters.

\section{Model independent sensitivity estimates on the anomalous quartic $WWWW$ couplings}

We examine the potential of $W^{+} W^{-}$ production at the future $ep$ colliders, namely at the LHeC and the FCC-he, on the anomalous $\frac{f_{M0}}{\Lambda^{4}}$, $\frac{f_{M1}}{\Lambda^{4}}$, $\frac{f_{M7}}{\Lambda^{4}}$, $\frac{f_{S0}}{\Lambda^{4}}$, $\frac{f_{S1}}{\Lambda^{4}}$, $\frac{f_{T0}}{\Lambda^{4}}$, $\frac{f_{T1}}{\Lambda^{4}}$ and $\frac{f_{T2}}{\Lambda^{4}}$ parameters of the anomalous quartic $WWWW$ couplings. To realize our work, we focus on the processes $ep \rightarrow \nu_{e}W^{+} W^{-}j$ with pure leptonic final state $ep \rightarrow \nu_{e}W^{+} W^{-}j\rightarrow \nu_{e} \ell^{+}\nu_{\ell}\ell^{-}\bar\nu_{\ell}j$ and semileptonic final state $ep \rightarrow \nu_{e}W^{+} W^{-}j\rightarrow \nu_{e} \ell^{+}\nu_{\ell}\ j j j$. In these processes, we consider that W's are on the mass shell, that is, we assumed it to be in the resonance region. We take into account the $e^{-}p \to e^{-}ZWj\to e^{-}\nu_{e}\bar{\nu_{e}} \mu^{+} \nu_{\mu} j$ process for the background in addition to the SM background for the leptonic decay channel. Also, $ep \to \nu W jjj \to \nu l \nu j j j $ process and SM background are considered for the semileptonic decay channel.

In Table X, we give the cross sections of $2\rightarrow6$ processes for both leptonic and semileptonic final state using by default cuts at $\sqrt{s}=10$ TeV. Comparing the signal processes, the most important contribution of the $2\rightarrow6$ process is generated through the signal processes. Also, we present the cross sections of the process $ep \to \nu_{e}W^{+}W^{-}j\rightarrow \nu_{e}\ell^{+}\nu_{\ell} j j j$ and the process $ep \to \nu_{e}Wj+ j j\to \nu_{e} \ell^{+}\nu_{\ell} j + j j $ for various center-of-mass energies in Table XI.

To study the sensitivity on the anomalous $\frac{f_{M0}}{\Lambda^{4}}$, $\frac{f_{M1}}{\Lambda^{4}}$, $\frac{f_{M7}}{\Lambda^{4}}$, $\frac{f_{S0}}{\Lambda^{4}}$, $\frac{f_{S1}}{\Lambda^{4}}$, $\frac{f_{T0}}{\Lambda^{4}}$, $\frac{f_{T1}}{\Lambda^{4}}$ and $\frac{f_{T2}}{\Lambda^{4}}$ parameters we apply $\chi^{2}$ test and investigate $95\%$ Confidence Level sensitivities.
The statistical method $\chi^{2}$ test is given by

\begin{eqnarray}
\chi^{2}=\left(\frac{\sigma_{SM}-\sigma_{NP}}{\sigma_{SM}\delta_{stat}}\right)^{2}
\end{eqnarray}
where $\delta_{stat}=\frac{1}{\sqrt{N_{SM}}}$ is the statistical error, $N_{SM}=L_{int}\times \sigma_{SM}$.

A significant subject in examing the anomalous quartic gauge boson couplings is to check for the preservation of unitarity. In effective Lagrangian approach, the cross sections of processes  including the anomalous gauge boson couplings increase with the coupling strength, and hence unitarity is violated for sufficiently high energy collisions.

Anomalous cross sections can grow rapidly with energy eventually violating the unitarity limit. To overcome this, an additional form factor is sometimes added to the coupling constant. Depending on the energy at which unitarity is violated, the form factors have an energy scale parameter $\Lambda$, which damps out the anomalous couplings. In the literature, while some studies on these anomalous quartic gauge couplings take this effect into account, some studies do not. In particular, in the last experimental article on the anomalous $WWWW$ coupling, the obtained limits on $\frac{f_{T0}}{\Lambda^{4}}$, $\frac{f_{T1}}{\Lambda^{4}}$ and $\frac{f_{T2}}{\Lambda^{4}}$ couplings were calculated without applying unitarization \cite{sir1}. As can be seen from the cross sections in Tables VI-IX, we do not consider violation of unitarity, when calculating the total cross sections. For example, the total cross section containing the $\frac{f_{T0}}{\Lambda^{4}}$ coupling is $10^{3}$ pb, whereas the SM cross section is $10^{-3}$ pb. Therefore, in our analysis, we do not apply any unitarity dipole form factor or cut off.

In Tables XII-XIX, constraints on the anomalous $\frac{f_{M0}}{\Lambda^{4}}$, $\frac{f_{M1}}{\Lambda^{4}}$, $\frac{f_{M7}}{\Lambda^{4}}$, $\frac{f_{S0}}{\Lambda^{4}}$, $\frac{f_{S1}}{\Lambda^{4}}$, $\frac{f_{T0}}{\Lambda^{4}}$, $\frac{f_{T1}}{\Lambda^{4}}$ and $\frac{f_{T2}}{\Lambda^{4}}$ parameters at 1.30, 1.98, 7.07 and 10 TeV future hadron electron colliders via $W^{+}W^{-}$ production pure leptonic and semileptonic decay channels with different integrated luminosities are shown. While obtaining the sensitivity limits for each coupling, all of the other couplings are set to zero at a time. We see from Tables XII-XIII that best sensitivities on $\frac{f_{S0}}{\Lambda^4}$ and $\frac{f_{S1}}{\Lambda^4}$ couplings are one order of magnitude worse than the experimental limits.  However, our limits for $\sqrt{s}=10$ TeV with $L_{int} = 1000$ fb$^{-1}$ from Table XV are comparable with the sensitivities of the LHC. In addition, our best limits on $\frac{f_{T0}}{\Lambda^{4}}$, $\frac{f_{T1}}{\Lambda^{4}}$ and $\frac{f_{T2}}{\Lambda^{4}}$ couplings are approximately one order of magnitude more stringent than the CMS Collaboration limits.

Furthermore, we compare our calculations with the phenomenological work at the LHC. In Ref. \cite{ebo}, the limits on the anomalous $\frac{f_{S0}}{\Lambda^4}$ and $\frac{f_{S1}}{\Lambda^4}$ couplings arising from dimension-8 effective Lagrangians are calculated as [-2.20; 2.40]$\times$ 10$^{1}$ TeV$^{-4}$ and [-2.50; 2.50]$\times$ 10$^{1}$ TeV$^{-4}$, respectively. As can be seen in Table XIX, our best limits on these couplings by probing the process $ep \rightarrow \nu_{e}W^{+} W^{-}j\rightarrow \nu_{e} \ell^{+}\nu_{\ell}\ j j j$ are more restrictive with respect to the limits obtained in Ref. \cite{ebo}. Also, the best limits on $\frac{f_{S1}}{\Lambda^4}$ coupling for pure leptonic decay channels of the $W^{+}W^{-}$ production in the final state of the process $ep \rightarrow \nu_{e}W^{+}W^{-} j$ at 10 TeV FCC-he with an integrated luminosity of 100 fb$^{-1}$ are up to four times better than the limits of Ref. \cite{ebo}.

In Snowmass paper \cite{deg}, $\frac{f_{T0}}{\Lambda^4}$ coupling at $95\%$ Confidence Level in pure leptonic channel via the triboson production is examined.  We observe that our best limit on this coupling at $\sqrt{s}=10$ TeV FCC-he with $L_{int} = 1000$ fb$^{-1}$ is almost twenty times worse than the best limits derived in Snowmass paper for $\sqrt{s}=100$ TeV with $L_{int} = 3000$ fb$^{-1}$.

Ref. \cite{yu} investigated the anomalous quartic couplings parameters $\frac{f_{S0}}{\Lambda^4}$, $\frac{f_{S1}}{\Lambda^4}$ and $\frac{f_{T0}}{\Lambda^4}$ at 100 TeV future $pp$ collider via $WWW$ production pure leptonic decay channel with integrated luminosity of 3000 fb$^{-1}$. They found $\frac{f_{S0}}{\Lambda^4}=[-2.93; 3.04]$ TeV$^{-4}$, $\frac{f_{S1}}{\Lambda^4}=[-1.30; 1.16]$ TeV$^{-4}$ and $\frac{f_{T0}}{\Lambda^4}=[-3.69; 2.97]\times 10^{-3}$ TeV$^{-4}$. As understood from Table XX, the limits on $\frac{f_{S0}}{\Lambda^4}$ and $\frac{f_{S1}}{\Lambda^4}$ couplings derived by Ref. \cite{yu} are very close to our best limits, but the best limits we find for $\frac{f_{T0}}{\Lambda^4}$ are about one order of magnitude worse.

In the literature, in addition to the $\frac{f_{S0}}{\Lambda^4}$, $\frac{f_{S1}}{\Lambda^4}$ and $\frac{f_{T0}}{\Lambda^4}$ parameters examined for the anomalous $WWWW$ couplings, we also investigate the anomalous $\frac{f_{M0}}{\Lambda^4}$, $\frac{f_{M1}}{\Lambda^4}$, $\frac{f_{M7}}{\Lambda^4}$, $\frac{f_{T1}}{\Lambda^4}$ and $\frac{f_{T2}}{\Lambda^4}$ parameters. The best limits on $\frac{f_{M0}}{\Lambda^4}$, $\frac{f_{M1}}{\Lambda^4}$, $\frac{f_{M7}}{\Lambda^4}$, $\frac{f_{T1}}{\Lambda^4}$ and $\frac{f_{T2}}{\Lambda^4}$ parameters
at the FCC-he with $\sqrt{s}=10$ TeV and $L_{int}=1000$ fb$^{-1}$ are obtained as [-1.06; 1.06] TeV$^{-4}$, [-4.10; 4.09] TeV$^{-1}$, [-7.67; 7.70] TeV$^{-4}$, [-2.55; 2.48]$\times$ 10$^{-1}$ TeV$^{-4}$ and [-4.40; 4.37]$\times$ 10$^{-1}$ TeV$^{-4}$, respectively.

Particularly, the sensitivity of the process $ep \rightarrow \nu_{e}W^{+}W^{-} j$ to the anomalous $\frac{f_{M0}}{\Lambda^4}$, $\frac{f_{M1}}{\Lambda^4}$, $\frac{f_{M7}}{\Lambda^4}$,$\frac{f_{S0}}{\Lambda^4}$, $\frac{f_{S1}}{\Lambda^4}$, $\frac{f_{T0}}{\Lambda^4}$, $\frac{f_{T1}}{\Lambda^4}$ and $\frac{f_{T2}}{\Lambda^4}$ parameters rapidly increases with the center-of-mass energy and the luminosity. As we can see from our results, we find that the process $ep \rightarrow \nu_{e}W^{+}W^{-} j$ is the most sensitive to $\frac{f_{T0}}{\Lambda^4}$ coupling that has stronger energy dependence than other couplings.

To make our study more effective, for $10$ TeV FCC-he with an integrated luminosity of $1000$ fb$^{-1}$, we find sensitivities on the anomalous $\frac{f_{M0}}{\Lambda^{4}}$, $\frac{f_{M1}}{\Lambda^{4}}$, $\frac{f_{M7}}{\Lambda^{4}}$, $\frac{f_{S0}}{\Lambda^{4}}$, $\frac{f_{S1}}{\Lambda^{4}}$, $\frac{f_{T0}}{\Lambda^{4}}$, $\frac{f_{T1}}{\Lambda^{4}}$ and $\frac{f_{T2}}{\Lambda^{4}}$ couplings by using statistical significance

\begin{eqnarray}
SS=\frac{|\sigma_{NP}-\sigma_{SM}|}{\sqrt{\sigma_{SM}}}\sqrt{L_{int}}.
\end{eqnarray}

For pure leptonic and semileptonic decay channels, $5(3)\sigma$ observation sensitivities on the anomalous couplings are given Tables XX-XXIII. For both decay channels, the obtained limits using $\chi^{2}$ analysis at $95\%$ Confidence Level are better than the best limits derived from signal significance at $3\sigma$ and $5\sigma$. Our best limits obtained on the anomalous couplings by using statistical significance at $3 \sigma$ are up to two times better than the best limits derived for $5 \sigma$.

There are other studies on the anomalous quartic $WWWW$ couplings through the nonlinear parametrization for the electroweak symmetry breaking sector in the literature \cite{11,12,baa}. We can compare the limits obtained from these studies with our own limits via the following relations \cite{baa}

\begin{eqnarray}
\alpha_{4}=\frac{f_{S0}}{\Lambda^4}\frac{\upsilon^{4}}{8},
\end{eqnarray}

\begin{eqnarray}
\alpha_{4}+2\alpha_{5}=\frac{f_{S1}}{\Lambda^4}\frac{\upsilon^{4}}{8}.
\end{eqnarray}

However, in this study, we will only focus on the researches that examine the anomalous quartic $WWWW$ couplings through the linear parametrization for the electroweak symmetry breaking sector.

\section{Conclusions}

The existence of self-interactions among the gauge bosons in EW sector is very useful to investigate the new physics effects coming from beyond the SM. Possible deviation of the quartic gauge boson couplings from the predictions of the SM would be a sign to new physics. The LHeC and the FCC-he with a rich physics program can provide a lot of important information to better understand the SM and to probe new physics. In this study, we offer an analysis to constrain the anomalous quartic gauge boson couplings in the electroweak sector by considering an effective Lagrangian approach with the content of several dimension-8 operators.

In this study, it is thought that both W bosons are produced on the mass shell. Tables XII-XV show the sensitivity limits on the anomalous quartic gauge couplings for the LHeC and FCC-he colliders for different center-of-mass energy values and for the pure leptonic decay channel. Similarly, sensitivity limits for semileptonic decay channels are given in Tables XVI-XIX. As can be seen from these tables, sensitivity limits imposed on pure leptonic and semileptonic decay channels differ at the LHeC and FCC-he colliders. The main reason for this phenomenon is that for the pure leptonic decay channel and the semileptonic decay channel, the cross sections for signal and background of the process have different trends for changing center-of-mass energy values. This effect can be best understood from the cross section differences that occur when cut parameters are applied and not applied. These results can be seen from Tables IV-V of efficiency values calculated from cross sections for leptonic and semi-leptonic decay of $W$-bosons at $\sqrt{s}=1.98$ TeV. The efficiency value is calculated as the ratio of the cross sections where the cuts are applied to the no-cut applied state. In Tables IV-V, the values of the anomalous couplings are taken as $1\times 10^{-9}$ GeV$^{-4}$. When tables are examined carefully, it can be seen that the efficiency value gives different responses in pure leptonic decays and semileptonic decays.

The best sensitivities obtained from the process $ep \rightarrow \nu_{e}W^{+} W^{-}j\rightarrow \nu_{e} \ell^{+}\nu_{\ell}\ell^{-}\nu_{\ell}j$ on the anomalous $\frac{f_{M0}}{\Lambda^{4}}$, $\frac{f_{M1}}{\Lambda^{4}}$, $\frac{f_{M7}}{\Lambda^{4}}$, $\frac{f_{S0}}{\Lambda^{4}}$, $\frac{f_{S1}}{\Lambda^{4}}$, $\frac{f_{T0}}{\Lambda^{4}}$, $\frac{f_{T1}}{\Lambda^{4}}$ and $\frac{f_{T2}}{\Lambda^{4}}$ couplings vary from the order of $10^{0}$ to the order of $10^{-2}$. In addition, the best sensitivities derived on the anomalous $\frac{f_{M0}}{\Lambda^{4}}$, $\frac{f_{M1}}{\Lambda^{4}}$, $\frac{f_{M7}}{\Lambda^{4}}$, $\frac{f_{S0}}{\Lambda^{4}}$, $\frac{f_{S1}}{\Lambda^{4}}$, $\frac{f_{T0}}{\Lambda^{4}}$, $\frac{f_{T1}}{\Lambda^{4}}$ and $\frac{f_{T2}}{\Lambda^{4}}$ couplings from the process $ep \rightarrow \nu_{e}W^{+} W^{-}j\rightarrow \nu_{e} \ell^{+}\nu_{\ell}\ j j j$ change from the order of $10^{1}$ to the order of $10^{-2}$.

We observe that the sensitivities obtained from the pure leptonic decay channel are better than those derived from the semileptonic channel. We have shown the sensitivity limits on the new physics parameters for pure leptonic decays in Tables XII-XV. When these tables are compared, it is seen that the sensitivity limits on new physics parameters improve in increasing center-of-mass energies. A similar situation for semileptonic decay channel can be seen when the Table XVI-XIX are examined. When tables are analysed for a single value of the center-of-mass energy, it is noticed that the sensitivity limits on the parameters become stronger with increasing luminosity values as expected.

$pp \rightarrow W W W$ process at the LHC is convenient to analyse the anomalous quartic
 $WWWW$  couplings. The $ep \rightarrow \nu_{e}W^{+} W^{-}j$ process 
provides a clean environment and lower background effects with respect to the $pp$ collision.
This creates a prominent adventage to the process $ep \rightarrow \nu_{e}W^{+} W^{-}j$ for the analysis of
isolated $WWWW$ quartic vertex.

Also, in this study, we give the first limits obtained on the anomalous $\frac{f_{M0}}{\Lambda^{4}}$, $\frac{f_{M1}}{\Lambda^{4}}$, $\frac{f_{M7}}{\Lambda^{4}}$, $\frac{f_{T1}}{\Lambda^{4}}$ and $\frac{f_{T2}}{\Lambda^{4}}$ couplings derived by dimension-8 operators at $ep$ colliders in the literature.

\pagebreak

\pagebreak

\begin{figure}
\includegraphics[width=7cm,height=4cm]{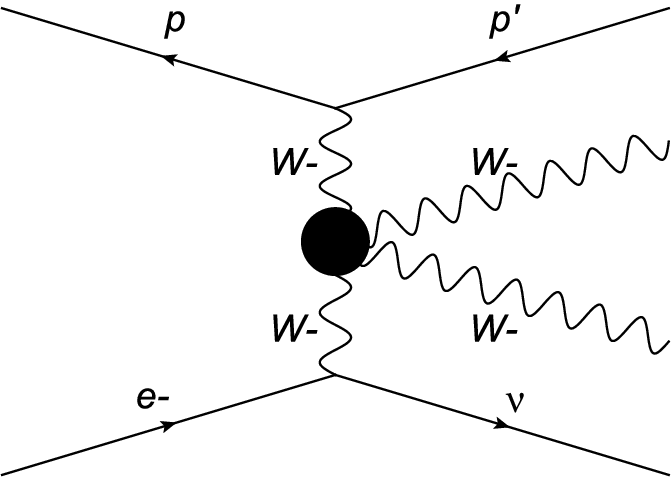}
\caption{Schematic diagram for the process $ep \rightarrow \nu_{e}W^{+}W^{-} j$.
\label{fig2}}
\end{figure}

\begin{figure}
\includegraphics{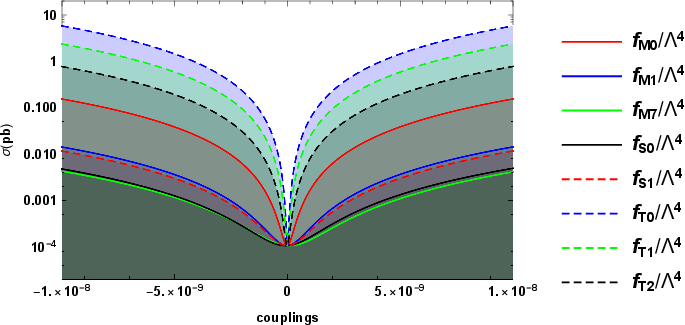}
\caption{For pure leptonic channel, the total cross sections dependences on the anomalous quartic $WWWW$ couplings $\frac{f_{M0}}{\Lambda^4}$, $\frac{f_{M1}}{\Lambda^4}$, $\frac{f_{M7}}{\Lambda^4}$, $\frac{f_{S0}}{\Lambda^4}$, $\frac{f_{S1}}{\Lambda^4}$, $\frac{f_{T0}}{\Lambda^4}$, $\frac{f_{T1}}{\Lambda^4}$ and $\frac{f_{T2}}{\Lambda^4}$ at 1.30 TeV LHeC. Here, the anomalous quartic couplings are in units of GeV$^{-4}$.
\label{fig2}}
\end{figure}

\begin{figure}
\includegraphics{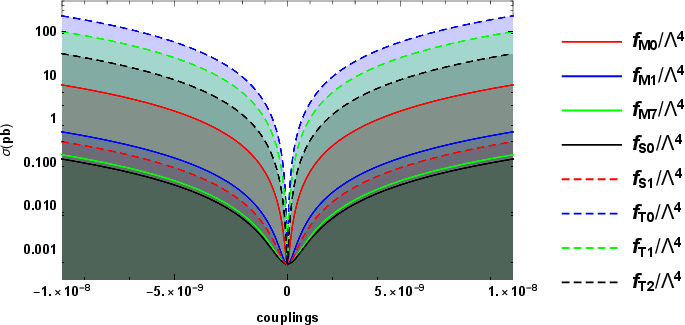}
\caption{The same as Fig. 2 but for 1.98 TeV LHeC.
\label{fig3}}
\end{figure}

\begin{figure}
\includegraphics{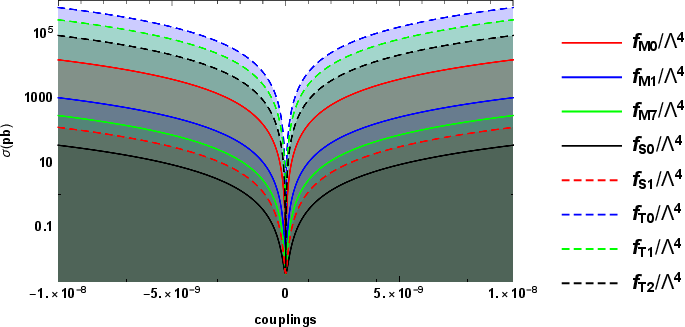}
\caption{For pure leptonic channel, the total cross sections dependences on the anomalous quartic $WWWW$ couplings $\frac{f_{M0}}{\Lambda^4}$, $\frac{f_{M1}}{\Lambda^4}$, $\frac{f_{M7}}{\Lambda^4}$, $\frac{f_{S0}}{\Lambda^4}$, $\frac{f_{S1}}{\Lambda^4}$, $\frac{f_{T0}}{\Lambda^4}$, $\frac{f_{T1}}{\Lambda^4}$ and $\frac{f_{T2}}{\Lambda^4}$ at 7.07 TeV FCC-he. Here, the anomalous quartic couplings are in units of GeV$^{-4}$.
\label{fig4}}
\end{figure}

\begin{figure}
\includegraphics{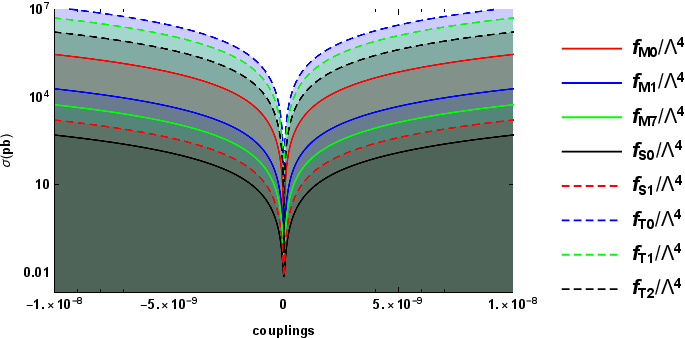}
\caption{The same as Fig. 4 but for 10 TeV FCC-he.
\label{fig5}}
\end{figure}

\begin{figure}
\includegraphics{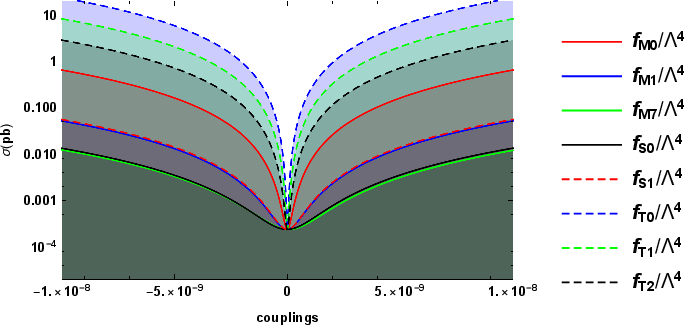}
\caption{For semileptonic channel, the total cross sections dependences on the anomalous quartic $WWWW$ couplings $\frac{f_{M0}}{\Lambda^4}$, $\frac{f_{M1}}{\Lambda^4}$, $\frac{f_{M7}}{\Lambda^4}$, $\frac{f_{S0}}{\Lambda^4}$, $\frac{f_{S1}}{\Lambda^4}$, $\frac{f_{T0}}{\Lambda^4}$, $\frac{f_{T1}}{\Lambda^4}$ and $\frac{f_{T2}}{\Lambda^4}$ at 1.30 TeV LHeC. Here, the anomalous quartic couplings are in units of GeV$^{-4}$.
\label{fig6}}
\end{figure}

\begin{figure}
\includegraphics{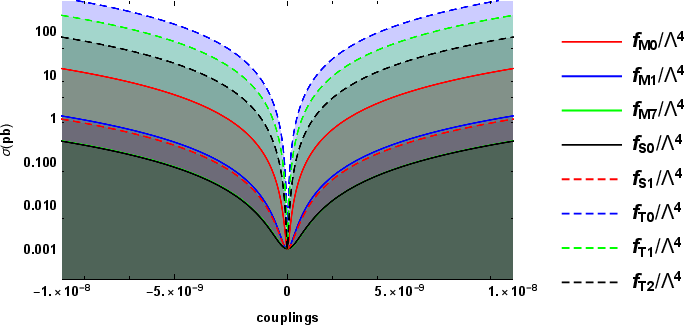}
\caption{The same as Fig. 6 but for 1.98 TeV LHeC.
\label{fig7}}
\end{figure}

\begin{figure}
\includegraphics{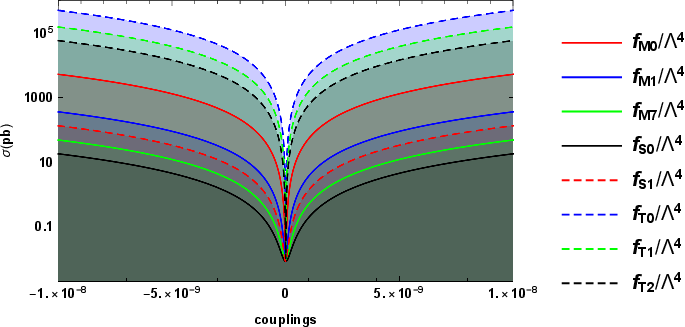}
\caption{For semileptonic channel, the total cross sections dependences on the anomalous quartic $WWWW$ couplings $\frac{f_{M0}}{\Lambda^4}$, $\frac{f_{M1}}{\Lambda^4}$, $\frac{f_{M7}}{\Lambda^4}$, $\frac{f_{S0}}{\Lambda^4}$, $\frac{f_{S1}}{\Lambda^4}$, $\frac{f_{T0}}{\Lambda^4}$, $\frac{f_{T1}}{\Lambda^4}$ and $\frac{f_{T2}}{\Lambda^4}$ at 7.07 TeV FCC-he. Here, the anomalous quartic couplings are in units of GeV$^{-4}$.
\label{fig8}}
\end{figure}

\begin{figure}
\includegraphics{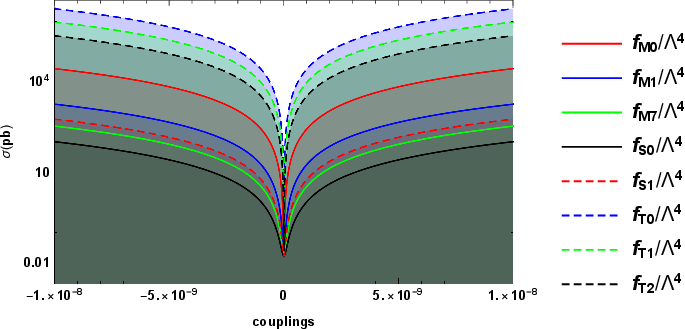}
\caption{The same as Fig. 8 but for 10 TeV FCC-he.
\label{fig9}}
\end{figure}

\begin{table}
\caption{The cross sections of signal for $10$ TeV$^{-4}$ value of anomalous $f_{M0}/\Lambda^{4}$, $f_{M1}/\Lambda^{4}$, $f_{M7}/\Lambda^{4}$, $f_{S0}/\Lambda^{4}$, $f_{S1}/\Lambda^{4}$, $f_{T0}/\Lambda^{4}$, $f_{T1}/\Lambda^{4}$ and $f_{T2}/\Lambda^{4}$ after cuts given in Eqs. 22-24 of the process $ep \rightarrow \nu_{e}W^{+} W^{-}j\rightarrow \nu_{e} \ell^{+}\nu_{\ell}\ell^{-}\nu_{\ell}j$ at the LHeC with $\sqrt{s}=1.30$ and $1.98$ TeV.
The SM cross sections of the process for $\sqrt{s}=1.30$ and $1.98$ TeV are $1.07\times 10^{-4}$ pb and $4.60\times 10^{-4}$ pb, respectively.}
\begin{tabular}{|c|c|c|}\hline
Couplings   & Signal at 1.30 TeV LHeC (pb) & Signal at 1.98 TeV LHeC (pb) \\
\hline \hline
$f_{M0}/\Lambda^{4}$  & 1.09 $\times 10^{-4}$ & 4.61 $\times 10^{-4}$ \\
\hline
$f_{M1}/\Lambda^{4}$  & 1.07 $\times 10^{-4}$ & 4.64 $ \times 10^{-4}$ \\
\hline
$f_{M7}/\Lambda^{4}$  & 1.08 $\times 10^{-4}$  & 4.59 $\times 10^{-4}$ \\
\hline
$f_{S0}/\Lambda^{4}$  &  1.09 $\times 10^{-4}$ & 4.61 $\times 10^{-4}$  \\
\hline
$f_{S1}/\Lambda^{4}$  &  1.09 $\times 10^{-4}$ & 4.59 $\times 10^{-4}$  \\
\hline
$f_{T0}/\Lambda^{4}$  &  1.20 $\times 10^{-4}$ & 7.19 $\times 10^{-4}$  \\
\hline
$f_{T1}/\Lambda^{4}$  & 1.15 $\times 10^{-4}$  & 5.81 $\times 10^{-4}$   \\
\hline
$f_{T2}/\Lambda^{4}$  & 1.11 $\times 10^{-4}$  & 5.14 $\times 10^{-4}$  \\
\hline
\end{tabular}
\end{table}

\begin{table}
\caption{The cross sections of signal for $1$ TeV$^{-4}$ value of anomalous $f_{M0}/\Lambda^{4}$, $f_{M1}/\Lambda^{4}$, $f_{M7}/\Lambda^{4}$, $f_{S0}/\Lambda^{4}$, $f_{S1}/\Lambda^{4}$, $f_{T0}/\Lambda^{4}$, $f_{T1}/\Lambda^{4}$ and $f_{T2}/\Lambda^{4}$ after cuts given in Eqs. 22-24 of the process $ep \rightarrow \nu_{e}W^{+} W^{-}j\rightarrow \nu_{e} \ell^{+}\nu_{\ell}\ell^{-}\nu_{\ell}j$ at the FCC-he with $\sqrt{s}=7.07$ and $10$ TeV.
The SM cross sections of the process for $\sqrt{s}=7.07$ and $10$ TeV are $3.60\times 10^{-3}$ pb and $7.35\times 10^{-3}$ pb, respectively.}
\begin{tabular}{|c|c|c|}\hline
Couplings   & Signal at 7.07 TeV FCC-he (pb) & Signal at 10 TeV FCC-he (pb) \\
\hline \hline
$f_{M0}/\Lambda^{4}$  & 3.73 $\times 10^{-3}$ & 9.62 $\times 10^{-3}$ \\
\hline
$f_{M1}/\Lambda^{4}$   & 3.72 $\times 10^{-3}$ & 7.63 $\times 10^{-3}$   \\
\hline
$f_{M7}/\Lambda^{4}$   & 3.70 $\times 10^{-3}$ & 7.42 $\times 10^{-3}$   \\
\hline
$f_{S0}/\Lambda^{4}$  & 3.64 $\times 10^{-3}$ & 7.44 $\times 10^{-3}$ \\
\hline
$f_{S1}/\Lambda^{4}$  & 3.69 $\times 10^{-3}$ & 7.29 $\times 10^{-3}$ \\
\hline
$f_{T0}/\Lambda^{4}$  & 9.73 $\times 10^{-3}$ & 1.23 $\times 10^{-1}$ \\
\hline
$f_{T1}/\Lambda^{4}$ & 6.20 $\times 10^{-3}$ & 5.62 $\times 10^{-2}$  \\
\hline
$f_{T2}/\Lambda^{4}$  & 4.40 $\times 10^{-3}$ & 2.09 $\times 10^{-2}$ \\
\hline
\end{tabular}
\end{table}

\begin{table}
\caption{The cross sections of signal for $10$ TeV$^{-4}$ value of anomalous $f_{M0}/\Lambda^{4}$, $f_{M1}/\Lambda^{4}$, $f_{M7}/\Lambda^{4}$, $f_{S0}/\Lambda^{4}$, $f_{S1}/\Lambda^{4}$, $f_{T0}/\Lambda^{4}$, $f_{T1}/\Lambda^{4}$ and $f_{T2}/\Lambda^{4}$ after cuts given in Eqs. 25-27 of the process $ep \rightarrow \nu_{e}W^{+} W^{-}j\rightarrow \nu_{e} \ell^{+}\nu_{\ell}\ j j j$ at the LHeC with $\sqrt{s}=1.30$ and $1.98$ TeV.
The SM cross sections of the process for $\sqrt{s}=1.30$ and $1.98$ TeV are $2.39\times 10^{-4}$ pb and $1.01\times 10^{-3}$ pb, respectively.}
\begin{tabular}{|c|c|c|}\hline
Couplings  & Signal at 1.30 TeV LHeC (pb) & Signal at 1.98 TeV LHeC (pb) \\
\hline \hline
$f_{M0}/\Lambda^{4}$  & 2.41 $\times 10^{-4}$ & 9.96 $\times 10^{-4}$\\
\hline
$f_{M1}/\Lambda^{4}$  & 2.43 $\times 10^{-4}$ & 1.00 $ \times 10^{-3}$ \\
\hline
$f_{M7}/\Lambda^{4}$  & 2.41 $\times 10^{-4}$  & 9.94 $\times 10^{-4}$ \\
\hline
$f_{S0}/\Lambda^{4}$  &  2.42 $\times 10^{-4}$ & 9.90 $\times 10^{-4}$  \\
\hline
$f_{S1}/\Lambda^{4}$  &  2.41 $\times 10^{-4}$ & 9.89 $\times 10^{-4}$  \\
\hline
$f_{T0}/\Lambda^{4}$  &  2.75 $\times 10^{-4}$ & 1.64 $\times 10^{-3}$  \\
\hline
$f_{T1}/\Lambda^{4}$  & 2.57 $\times 10^{-4}$  & 1.26 $\times 10^{-3}$  \\
\hline
$f_{T2}/\Lambda^{4}$  & 2.52 $\times 10^{-4}$  & 1.13 $\times 10^{-3}$  \\
\hline
\end{tabular}
\end{table}

\begin{table}
\caption{The cross sections of signal for $1$ TeV$^{-4}$ value of anomalous $f_{M0}/\Lambda^{4}$, $f_{M1}/\Lambda^{4}$, $f_{M7}/\Lambda^{4}$, $f_{S0}/\Lambda^{4}$, $f_{S1}/\Lambda^{4}$, $f_{T0}/\Lambda^{4}$, $f_{T1}/\Lambda^{4}$ and $f_{T2}/\Lambda^{4}$ after cuts given in Eqs. 25-27 of the process $ep \rightarrow \nu_{e}W^{+} W^{-}j\rightarrow \nu_{e} \ell^{+}\nu_{\ell}\ j j j$ at the FCC-he with $\sqrt{s}=7.07$ and $10$ TeV.
The SM cross sections of the process for $\sqrt{s}=7.07$ and $10$ TeV are $8.41\times 10^{-3}$ pb and $1.62\times 10^{-2}$ pb, respectively.}
\begin{tabular}{|c|c|c|}\hline
Couplings   & Signal at 7.07 TeV FCC-he (pb) & Signal at 10 TeV FCC-he (pb) \\
\hline \hline
$f_{M0}/\Lambda^{4}$  & 8.44 $\times 10^{-3}$ & 1.67 $\times 10^{-2}$ \\
\hline
$f_{M1}/\Lambda^{4}$   & 8.28 $\times 10^{-3}$  & 1.67 $\times 10^{-2}$   \\
\hline
$f_{M7}/\Lambda^{4}$   & 8.48 $\times 10^{-3}$ & 1.63 $\times 10^{-2}$   \\
\hline
$f_{S0}/\Lambda^{4}$  & 8.39 $\times 10^{-3}$ & 1.65 $\times 10^{-2}$ \\
\hline
$f_{S1}/\Lambda^{4}$  & 8.39 $\times 10^{-3}$ & 1.65 $\times 10^{-2}$ \\
\hline
$f_{T0}/\Lambda^{4}$  & 1.21 $\times 10^{-2}$ & 3.30 $\times 10^{-2}$ \\
\hline
$f_{T1}/\Lambda^{4}$ & 9.55 $\times 10^{-3}$ & 2.33 $\times 10^{-2}$  \\
\hline
$f_{T2}/\Lambda^{4}$  & 8.95 $\times 10^{-3}$ & 1.75 $\times 10^{-2}$ \\
\hline
\end{tabular}
\end{table}

\begin{table}
\begin{center}
\caption{The cross sections of the processes $2\rightarrow 4$ and $2\rightarrow 6$. }
\begin{tabular}{|c|c|c|c|c|c|}\hline
  & $ep \to \nu_{e}W^{+}(W^{+} \rightarrow \ell^{+}\nu_{\ell})W^{-}(W^{-} \rightarrow \ell^{-}\bar\nu_{\ell}) j$ & $ep \to \nu_{e} \ell^{+}\nu_{\ell}  \ell^{-}\bar\nu_{\ell} j$  \\
\hline \hline
$\sqrt{s}=10$ TeV & $7.35\times 10^{-3}$ pb & $1.84\times 10^{-2}$ pb  \\
\hline \hline
&$ep \to \nu_{e}W^{+}(W^{+} \rightarrow \ell^{+}\nu_{\ell})W^{-}(W^{-} \rightarrow j j )j$ & $ep \to \nu_{e} \ell^{+}\nu_{\ell} j j j$   \\
\hline \hline
$\sqrt{s}=10$ TeV &  $1.62\times 10^{-2}$ pb &  $2.70\times 10^{-2}$ pb \\
\hline
\end{tabular}
\end{center}
\end{table}

\begin{table}
\begin{center}
\caption{The cross sections of the processes }
\begin{tabular}{|c|c|c|c|c|c|}\hline
  & $ep \to \nu_{e}W^{+}(W^{+} \rightarrow \ell^{+}\nu_{\ell})W^{-}(W^{-} \rightarrow j j )j$ & $ep \to \nu_{e}W(W \rightarrow \ell^{+}\nu_{\ell}) j+ j j$   \\
\hline \hline
$\sqrt{s}=1.30$ TeV &  $2.39\times 10^{-4}$ pb &  $8.53\times 10^{-6}$ pb \\
\hline
$\sqrt{s}=1.98$ TeV & $1.01\times 10^{-3}$ pb & $3.13\times 10^{-5}$ pb   \\
\hline
$\sqrt{s}=7.07$ TeV & $8.41\times 10^{-3}$ pb & $4.34\times 10^{-4}$ pb   \\
\hline
$\sqrt{s}=10$ TeV & $1.62\times 10^{-2}$ pb & $7.65\times 10^{-4}$ pb  \\
\hline
\end{tabular}
\end{center}
\end{table}

\begin{table}
\caption{Limits on the anomalous $\frac{f_{M0}}{\Lambda^{4}}$, $\frac{f_{M1}}{\Lambda^{4}}$, $\frac{f_{M7}}{\Lambda^{4}}$, $\frac{f_{S0}}{\Lambda^{4}}$, $\frac{f_{S1}}{\Lambda^{4}}$, $\frac{f_{T0}}{\Lambda^{4}}$, $\frac{f_{T1}}{\Lambda^{4}}$ and $\frac{f_{T2}}{\Lambda^{4}}$ couplings at 1.30 TeV LHeC through $W^{+}W^{-}$ production pure leptonic decay channel with integrated luminosities of 10, 30, 50 and 100 fb$^{-1}$. Here, the obtained anomalous quartic couplings are in units of TeV$^{-4}$.}
\begin{tabular}{|c|c|c|c|c|c|}\hline
Couplings  & 10 fb$^{-1}$ & 30 fb$^{-1}$ & 50 fb$^{-1}$ & 100 fb$^{-1}$ \\
\hline \hline
$f_{M0}/\Lambda^{4}$  & [-4.50;4.39] $\times 10^{2}$ & [-3.65;3.51] $\times 10^{2}$ & [-3.34;3.21] $\times 10^{2}$ & [-3.00;2.87] $\times 10^{2}$ \\
\hline
$f_{M1}/\Lambda^{4}$  & [-1.47;1.45] $\times 10^{3}$ & [-1.19;1.17]$ \times 10^{3}$  & [-1.09;1.07] $ \times 10^{3}$ & [-0.98;0.95]$\times 10^{3}$   \\
\hline
$f_{M7}/\Lambda^{4}$  & [-2.63;2.80]$\times 10^{3}$    & [-2.11;2.27]$\times 10^{3}$   & [-1.92;2.09]$\times 10^{3}$   & [-1.71;1.88]$\times 10^{3}$    \\
\hline
$f_{S0}/\Lambda^{4}$  &  [-2.60;2.44] $\times 10^{3}$ & [-2.11;1.95] $\times 10^{3}$ & [-1.94;1.78] $\times 10^{3}$ & [-1.75;1.59] $\times 10^{3}$ \\
\hline
$f_{S1}/\Lambda^{4}$  & [-1.62;1.61] $\times 10^{3}$  & [-1.31;1.30] $\times 10^{3}$ & [-1.20;1.19] $\times 10^{3}$ & [-1.07;1.06] $\times 10^{3}$  \\
\hline
$f_{T0}/\Lambda^{4}$  &  [-7.03;7.49] $\times 10^{1}$ & [-5.64;6.09] $\times 10^{1}$ & [-5.13;5.58] $\times 10^{1}$ & [-4.57;5.02] $\times 10^{1}$ \\
\hline
$f_{T1}/\Lambda^{4}$  & [-3.63;3.56] $\times 10^{1}$  & [-2.93;2.87] $\times 10^{1}$ & [-2.68;2.62] $\times 10^{1}$ & [-2.41;2.34] $\times 10^{1}$  \\
\hline
$f_{T2}/\Lambda^{4}$  & [-2.10;1.87] $\times 10^{2}$  & [-1.72;1.49] $\times 10^{2}$ & [-1.58;1.35] $\times 10^{2}$ & [-1.43;1.20] $\times 10^{2}$ \\
\hline
\end{tabular}
\end{table}

\begin{table}
\caption{The same as Table XII but for 1.98 TeV LHeC.}
\begin{tabular}{|c|c|c|c|c|c|}\hline
Couplings  & 10 fb$^{-1}$ & 30 fb$^{-1}$ & 50 fb$^{-1}$ & 100 fb$^{-1}$ \\
\hline \hline
$f_{M0}/\Lambda^{4}$  & [-1.08;1.07] $\times 10^{2}$ & [-8.98;8.94] $\times 10^{1}$ & [-8.35;8.31] $\times 10^{1}$ & [-7.66;7.62] $\times 10^{1}$ \\
\hline
$f_{M1}/\Lambda^{4}$  & [-3.72;3.72] $\times 10^{2}$ & [-3.10;3.10]$ \times 10^{2}$  & [-2.88;2.88] $ \times 10^{2}$ & [-2.65;2.64]$\times 10^{2}$   \\
\hline
$f_{M7}/\Lambda^{4}$  & [-6.81;6.85]$\times 10^{2}$    & [-5.67;5.71]$\times 10^{2}$   & [-5.27;5.31]$\times 10^{2}$   & [-4.83;4.87]$\times 10^{2}$    \\
\hline
$f_{S0}/\Lambda^{4}$  &  [-7.69;7.61] $\times 10^{2}$ & [-6.41;6.33] $\times 10^{2}$ & [-5.96;5.88] $\times 10^{2}$ & [-5.47;5.39] $\times 10^{2}$ \\
\hline
$f_{S1}/\Lambda^{4}$  & [-4.83;4.82] $\times 10^{2}$  & [-4.02;4.02] $\times 10^{2}$ & [-3.74;3.73] $\times 10^{2}$ & [-3.43;3.43] $\times 10^{2}$  \\
\hline
$f_{T0}/\Lambda^{4}$  &  [-1.72;1.76] $\times 10^{1}$ & [-1.43;1.47] $\times 10^{1}$ & [-1.32;1.36]$\times 10^{1}$ & [-1.21;1.25]$\times 10^{1}$ \\
\hline
$f_{T1}/\Lambda^{4}$  & [-2.70;2.56] $\times 10^{1}$  & [-2.26;2.12] $\times 10^{1}$ & [-2.10;1.97] $\times 10^{1}$ & [-1.94;1.80] $\times 10^{1}$  \\
\hline
$f_{T2}/\Lambda^{4}$  & [-4.73;4.69] $\times 10^{1}$  & [-3.95;3.90] $\times 10^{1}$ & [-3.67;3.62] $\times 10^{1}$ & [-3.37;3.32] $\times 10^{1}$ \\
\hline
\end{tabular}
\end{table}

\begin{table}
\caption{Limits on the anomalous $\frac{f_{M0}}{\Lambda^{4}}$, $\frac{f_{M1}}{\Lambda^{4}}$, $\frac{f_{M7}}{\Lambda^{4}}$, $\frac{f_{S0}}{\Lambda^{4}}$, $\frac{f_{S1}}{\Lambda^{4}}$, $\frac{f_{T0}}{\Lambda^{4}}$, $\frac{f_{T1}}{\Lambda^{4}}$ and $\frac{f_{T2}}{\Lambda^{4}}$ couplings at 7.07 TeV FCC-he through $W^{+}W^{-}$ production pure leptonic decay channel with integrated luminosities of 100, 300, 500 and 1000 fb$^{-1}$. Here, the obtained anomalous quartic couplings are in units of TeV$^{-4}$.}
\begin{tabular}{|c|c|c|c|c|c|}\hline
Couplings  & 100 fb$^{-1}$ & 300 fb$^{-1}$ & 500 fb$^{-1}$ & 1000 fb$^{-1}$ \\
\hline \hline
$f_{M0}/\Lambda^{4}$  & [-4.14;4.13] & [-3.98;3.96] & [-3.93;3.91] & [-3.88;3.86]  \\
\hline
$f_{M1}/\Lambda^{4}$  & [-1.60;1.59]$\times 10^{1}$ & [-1.53;1.53]$\times 10^{1}$  & [-1.51;1.51]$\times 10^{1}$ & [-1.49;1.49]$\times 10^{1}$   \\
\hline
$f_{M7}/\Lambda^{4}$  & [-2.99;3.03]$\times 10^{1}$    & [-2.87;2.91]$\times 10^{1}$  & [-2.83;2.88]$\times 10^{1}$  & [-2.79;2.84]$\times 10^{1}$    \\
\hline
$f_{S0}/\Lambda^{4}$  &  [-8.69;8.67] $\times 10^{1}$ & [-8.34;8.32] $\times 10^{1}$ & [-8.23;8.22] $\times 10^{1}$ & [-8.12;8.10] $\times 10^{1}$ \\
\hline
$f_{S1}/\Lambda^{4}$  & [-4.63;4.56] $\times 10^{1}$  & [-4.44;4.38] $\times 10^{1}$ & [-4.39;4.32] $\times 10^{1}$ & [-4.33;4.26] $\times 10^{1}$ \\
\hline
$f_{T0}/\Lambda^{4}$  &  [-5.96;6.91] $\times 10^{-1}$ & [-5.71;6.66] $\times 10^{-1}$ & [-5.63;6.58] $\times 10^{-1}$ & [-5.55;6.50] $\times 10^{-1}$ \\
\hline
$f_{T1}/\Lambda^{4}$  & [-9.98;9.90]$\times 10^{-1}$ & [-9.59;9.51]$\times 10^{-1}$  & [-9.46;9.38]$\times 10^{-1}$ & [-9.33;9.26]$\times 10^{-1}$   \\
\hline
$f_{T2}/\Lambda^{4}$  & [-1.77;1.71] & [-1.70;1.64] & [-1.68;1.62]  & [-1.65;1.60]  \\
\hline
\end{tabular}
\end{table}

\begin{table}
\caption{The same as Table XIV but for 10 TeV FCC-he.}
\begin{tabular}{|c|c|c|c|c|c|}\hline
Couplings  & 100 fb$^{-1}$ & 300 fb$^{-1}$ & 500 fb$^{-1}$ & 1000 fb$^{-1}$ \\
\hline \hline
$f_{M0}/\Lambda^{4}$  & [-1.13;1.13] & [-1.09;1.08]  & [-1.07;1.07] & [-1.06;1.06]  \\
\hline
$f_{M1}/\Lambda^{4}$  & [-4.37;4.36] & [-4.21;4.20]  & [-4.15;4.14] & [-4.10;4.09]   \\
\hline
$f_{M7}/\Lambda^{4}$  & [-8.18;8.20]    & [-7.87;7.89]    & [-7.77;7.80]  & [-7.67;7.70]     \\
\hline
$f_{S0}/\Lambda^{4}$  &  [-2.72;2.68] $\times 10^{1}$ & [-2.62;2.58] $\times 10^{1}$ & [-2.58;2.55] $\times 10^{1}$ & [-2.55;2.52] $\times 10^{1}$\\
\hline
$f_{S1}/\Lambda^{4}$  & [-1.50;1.49] $\times 10^{1}$  & [-1.45;1.43]$\times 10^{1}$ & [-1.43;1.41]$\times 10^{1}$ & [-1.41;1.40]$\times 10^{1}$  \\
\hline
$f_{T0}/\Lambda^{4}$  &  [-1.76;1.78] $\times 10^{-1}$ & [-1.69;1.72] $\times 10^{-1}$ & [-1.67;1.69] $\times 10^{-1}$ & [-1.65;1.67] $\times 10^{-1}$ \\
\hline
$f_{T1}/\Lambda^{4}$  & [-2.71;2.64] $\times 10^{-1}$  & [-2.61;2.54] $\times 10^{-1}$ & [-2.58;2.51] $\times 10^{-1}$ & [-2.55;2.48] $\times 10^{-1}$  \\
\hline
$f_{T2}/\Lambda^{4}$  & [-4.69;4.66] $\times 10^{-1}$  & [-4.51;4.48] $\times 10^{-1}$ & [-4.45;4.43] $\times 10^{-1}$ & [-4.40;4.37] $\times 10^{-1}$ \\
\hline
\end{tabular}
\end{table}

\begin{table}
\caption{Limits on the anomalous $\frac{f_{M0}}{\Lambda^{4}}$, $\frac{f_{M1}}{\Lambda^{4}}$, $\frac{f_{M7}}{\Lambda^{4}}$, $\frac{f_{S0}}{\Lambda^{4}}$, $\frac{f_{S1}}{\Lambda^{4}}$, $\frac{f_{T0}}{\Lambda^{4}}$, $\frac{f_{T1}}{\Lambda^{4}}$ and $\frac{f_{T2}}{\Lambda^{4}}$ couplings at 1.30 TeV LHeC through $W^{+}W^{-}$ production semileptonic decay channel with integrated luminosities of 10, 30, 50 and 100 fb$^{-1}$. Here, the obtained anomalous quartic couplings are in units of TeV$^{-4}$.}
\begin{tabular}{|c|c|c|c|c|c|}\hline
Couplings  & 10 fb$^{-1}$ & 30 fb$^{-1}$ & 50 fb$^{-1}$ & 100 fb$^{-1}$ \\
\hline \hline
$f_{M0}/\Lambda^{4}$  & [-2.23;2.21] $\times 10^{2}$ & [-1.71;1.70] $\times 10^{2}$ & [-1.52;1.50] $\times 10^{2}$ & [-1.29;1.28] $\times 10^{2}$ \\
\hline
$f_{M1}/\Lambda^{4}$  & [-7.99;7.71] $\times 10^{2}$ & [-6.17;5.88]$ \times 10^{2}$  & [-5.48;5.20] $ \times 10^{2}$ & [-4.69;4.40]$\times 10^{2}$   \\
\hline
$f_{M7}/\Lambda^{4}$  & [-1.57;1.70]$\times 10^{3}$    & [-1.19;1.32]$\times 10^{3}$   & [-1.05;1.17]$\times 10^{3}$   & [-8.84;1.01]$\times 10^{3}$    \\
\hline
$f_{S0}/\Lambda^{4}$  &  [-1.58;1.53] $\times 10^{3}$ & [-1.22;1.17] $\times 10^{3}$ & [-1.08;1.04] $\times 10^{3}$ & [-9.21;8.79] $\times 10^{2}$ \\
\hline
$f_{S1}/\Lambda^{4}$  & [-7.63;7.53] $\times 10^{2}$  & [-5.87;5.77] $\times 10^{2}$ & [-5.21;5.11] $\times 10^{2}$ & [-4.44;4.34] $\times 10^{2}$  \\
\hline
$f_{T0}/\Lambda^{4}$  &  [-3.64;3.77] $\times 10^{1}$ & [-2.77;2.91] $\times 10^{1}$ & [-2.45;2.59] $\times 10^{1}$ & [-2.08;2.21] $\times 10^{1}$ \\
\hline
$f_{T1}/\Lambda^{4}$  & [-6.42;6.08] $\times 10^{1}$  & [-4.96;4.63] $\times 10^{1}$ & [-4.42;4.08] $\times 10^{1}$ & [-3.78;3.45] $\times 10^{1}$  \\
\hline
$f_{T2}/\Lambda^{4}$  & [-1.12;1.00] $\times 10^{2}$  & [-8.72;7.56] $\times 10^{1}$ & [-7.80;6.64] $\times 10^{1}$ & [-6.73;5.57] $\times 10^{1}$ \\
\hline
\end{tabular}
\end{table}

\begin{table}
\caption{The same as Table XVI but for 1.98 TeV LHeC.}
\begin{tabular}{|c|c|c|c|c|c|}\hline
Couplings  & 10 fb$^{-1}$ & 30 fb$^{-1}$ & 50 fb$^{-1}$ & 100 fb$^{-1}$ \\
\hline \hline
$f_{M0}/\Lambda^{4}$  & [-7.04;6.83] $\times 10^{1}$ & [-5.46;5.26] $\times 10^{1}$ & [-4.87;4.67] $\times 10^{1}$ & [-4.20;3.99] $\times 10^{1}$ \\
\hline
$f_{M1}/\Lambda^{4}$  & [-2.51;2.34] $\times 10^{2}$ & [-1.96;1.79]$ \times 10^{2}$  & [-1.75;1.59] $ \times 10^{2}$ & [-1.52;1.35]$\times 10^{2}$   \\
\hline
$f_{M7}/\Lambda^{4}$  & [-4.54;4.85]$\times 10^{2}$    & [-3.48;3.79]$\times 10^{2}$   & [-3.08;3.39]$\times 10^{2}$   & [-2.62;2.93]$\times 10^{2}$    \\
\hline
$f_{S0}/\Lambda^{4}$  &  [-4.76;4.72] $\times 10^{2}$ & [-3.70;3.65] $\times 10^{2}$ & [-3.27;3.25] $\times 10^{2}$ & [-2.81;2.78] $\times 10^{2}$ \\
\hline
$f_{S1}/\Lambda^{4}$  & [-2.66;2.62] $\times 10^{2}$  & [-2.06;2.02] $\times 10^{2}$ & [-1.83;1.80] $\times 10^{2}$ & [-1.58;1.54] $\times 10^{2}$  \\
\hline
$f_{T0}/\Lambda^{4}$  &  [-1.05;1.10] $\times 10^{1}$ & [-8.10;8.52] & [-7.19;7.61] & [-6.14;6.56] \\
\hline
$f_{T1}/\Lambda^{4}$  & [-1.70;1.69] $\times 10^{1}$  & [-1.32;1.31] $\times 10^{1}$ & [-1.17;1.16] $\times 10^{1}$ & [-1.01;1.00] $\times 10^{1}$ \\
\hline
$f_{T2}/\Lambda^{4}$  & [-3.04;2.99] $\times 10^{1}$  & [-2.36;2.31] $\times 10^{1}$ & [-2.10;2.05] $\times 10^{1}$ & [-1.81;1.76] $\times 10^{1}$ \\
\hline
\end{tabular}
\end{table}

\begin{table}
\caption{Limits on the anomalous $\frac{f_{M0}}{\Lambda^{4}}$, $\frac{f_{M1}}{\Lambda^{4}}$, $\frac{f_{M7}}{\Lambda^{4}}$, $\frac{f_{S0}}{\Lambda^{4}}$, $\frac{f_{S1}}{\Lambda^{4}}$, $\frac{f_{T0}}{\Lambda^{4}}$, $\frac{f_{T1}}{\Lambda^{4}}$ and $\frac{f_{T2}}{\Lambda^{4}}$ couplings at 7.07 TeV FCC-he through $W^{+}W^{-}$ production semileptonic decay channel with integrated luminosities of 100, 300, 500 and 1000 fb$^{-1}$. Here, the obtained anomalous quartic couplings are in units of TeV$^{-4}$.}
\begin{tabular}{|c|c|c|c|c|c|}\hline
Couplings  & 100 fb$^{-1}$ & 300 fb$^{-1}$ & 500 fb$^{-1}$ & 1000 fb$^{-1}$ \\
\hline \hline
$f_{M0}/\Lambda^{4}$  & [-4.52;4.32] & [-3.95;3.75] & [-3.76;3.55] & [-3.55;3.34] \\
\hline
$f_{M1}/\Lambda^{4}$  & [-1.69;1.67] $\times 10^{1}$ & [-1.48;1.45] $\times 10^{1}$ & [1.40;1.38] $\times 10^{1}$ & [-1.32;1.30] $\times 10^{1}$  \\
\hline
$f_{M7}/\Lambda^{4}$  & [-4.59;4.60]$\times 10^{1}$    & [-3.99;4.01] $\times 10^{1}$ & [-3.79;3.80] $\times 10^{1}$ & [-3.58;3.59] $\times 10^{1}$   \\
\hline
$f_{S0}/\Lambda^{4}$  &  [-7.54;7.48] $\times 10^{1}$ & [-6.57;6.51] $\times 10^{1}$ & [-6.24;6.18] $\times 10^{1}$ & [-5.89;5.83] $\times 10^{1}$ \\
\hline
$f_{S1}/\Lambda^{4}$  & [-2.78;2.75] $\times 10^{1}$  & [-2.42;2.39] $\times 10^{1}$ & [-2.30;2.27] $\times 10^{1}$ & [-2.17;2.14] $\times 10^{1}$  \\
\hline
$f_{T0}/\Lambda^{4}$  &  [-4.37;4.71] $\times 10^{-1}$ & [-3.79;4.13] $\times 10^{-1}$ & [-3.59;3.93] $\times 10^{-1}$ & [-3.37;3.71] $\times 10^{-1}$ \\
\hline
$f_{T1}/\Lambda^{4}$  & [-8.21;8.17] $\times 10^{-1}$  & [-7.15;7.11] $\times 10^{-1}$ & [-6.79;6.75] $\times 10^{-1}$ & [-6.40;6.37] $\times 10^{-1}$  \\
\hline
$f_{T2}/\Lambda^{4}$  & [-1.37;1.29]  & [-1.20;1.12] & [-1.14;1.06] & [-1.07;1.00] \\
\hline
\end{tabular}
\end{table}

\begin{table}
\caption{The same as Table XVIII but for 10 TeV FCC-he.}
\begin{tabular}{|c|c|c|c|c|c|}\hline
Couplings  & 100 fb$^{-1}$ & 300 fb$^{-1}$ & 500 fb$^{-1}$ & 1000 fb$^{-1}$ \\
\hline \hline
$f_{M0}/\Lambda^{4}$  & [-2.51;2.33] & [-2.23;2.04] & [-2.14;1.95] & [-2.04;1.85] \\
\hline
$f_{M1}/\Lambda^{4}$  & [-9.49;9.26] & [-8.40;8.17]  & [-8.04;7.81] & [-7.65;7.42]   \\
\hline
$f_{M7}/\Lambda^{4}$  & [-2.16;2.18]  $\times 10^{1}$  & [-1.91;1.93] $\times 10^{1}$  & [-1.82;1.84] $\times 10^{1}$  & [-1.73;1.75] $\times 10^{1}$   \\
\hline
$f_{S0}/\Lambda^{4}$  &  [-3.96;3.90] $\times 10^{1}$ & [-3.50;3.44] $\times 10^{1}$ & [-3.35;3.29] $\times 10^{1}$ & [-3.19;3.13] $\times 10^{1}$ \\
\hline
$f_{S1}/\Lambda^{4}$  & [-1.67;1.66] $\times 10^{1}$  & [-1.47;1.47] $\times 10^{1}$ & [-1.41;1.40] $\times 10^{1}$ & [-1.34;1.33] $\times 10^{1}$  \\
\hline
$f_{T0}/\Lambda^{4}$  &  [-2.46;2.46] $\times 10^{-1}$ & [-2.17;2.17] $\times 10^{-1}$ & [-2.07;2.08] $\times 10^{-1}$ & [-1.97;1.98] $\times 10^{-1}$ \\
\hline
$f_{T1}/\Lambda^{4}$  & [-4.10;4.05] $\times 10^{-1}$  & [-3.62;3.58] $\times 10^{-1}$ & [-3.46;3.42] $\times 10^{-1}$ & [-3.30;3.25] $\times 10^{-1}$  \\
\hline
$f_{T2}/\Lambda^{4}$  & [-6.93;6.86] $\times 10^{-1}$ & [-6.13;6.06] $\times 10^{-1}$ & [-5.86;5.79] $\times 10^{-1}$ & [-5.58;5.50] $\times 10^{-1}$ \\
\hline
\end{tabular}
\end{table}

\begin{table}
\caption{For 10 TeV FCC-he with integrated luminosities of 100, 300, 500 and 1000 fb$^{-1}$ through $W^{+}W^{-}$ production pure leptonic decay channel, limits on the anomalous $\frac{f_{M0}}{\Lambda^{4}}$, $\frac{f_{M1}}{\Lambda^{4}}$, $\frac{f_{M7}}{\Lambda^{4}}$, $\frac{f_{S0}}{\Lambda^{4}}$, $\frac{f_{S1}}{\Lambda^{4}}$, $\frac{f_{T0}}{\Lambda^{4}}$, $\frac{f_{T1}}{\Lambda^{4}}$ and $\frac{f_{T2}}{\Lambda^{4}}$ couplings by using statistical significance at $3 \sigma$. Here, the obtained anomalous quartic couplings are in units of TeV$^{-4}$.}
\begin{tabular}{|c|c|c|c|c|c|}\hline
Couplings (TeV$^{-4}$) & 100 fb$^{-1}$ & 300 fb$^{-1}$ & 500 fb$^{-1}$ & 1000 fb$^{-1}$ \\
\hline \hline
$f_{M0}/\Lambda^{4}$  & [-1.18;1.18]  & [-1.12;1.11]  & [-1.10;1.09]  & [-1.08;1.07]  \\
\hline
$f_{M1}/\Lambda^{4}$  & [-4.57;4.56] & [-4.33;4.32]  & [-4.25;4.24] & [-4.17;4.16]   \\
\hline
$f_{M7}/\Lambda^{4}$  & [-8.55;8.57]   & [-8.09;8.11]  & [-7.95;7.97]  & [-7.80;7.82]    \\
\hline
$f_{S0}/\Lambda^{4}$  &  [-2.84;2.80] $\times 10^{1}$ & [-2.69;2.65]$\times 10^{1}$ & [-2.64;2.61]$\times 10^{1}$ & [-2.59;2.56]$\times 10^{1}$  \\
\hline
$f_{S1}/\Lambda^{4}$  & [-1.57;1.56]$\times 10^{1}$  & [-1.49;1.47]$\times 10^{1}$ & [-1.46;1.45]$\times 10^{1}$ & [-1.43;1.42]$\times 10^{1}$  \\
\hline
$f_{T0}/\Lambda^{4}$  &  [-1.84;1.86] $\times 10^{-1}$ & [-1.74;1.76] $\times 10^{-1}$ & [-1.71;1.73] $\times 10^{-1}$ & [-1.68;1.70] $\times 10^{-1}$ \\
\hline
$f_{T1}/\Lambda^{4}$  & [-2.84;2.76] $\times 10^{-1}$  & [-2.69;2.62] $\times 10^{-1}$ & [-2.64;2.57] $\times 10^{-1}$ & [-2.59;2.52] $\times 10^{-1}$  \\
\hline
$f_{T2}/\Lambda^{4}$  & [-4.90;4.87] $\times 10^{-1}$ & [-4.64;4.61] $\times 10^{-1}$ & [-4.56;4.53] $\times 10^{-1}$ & [-4.47;4.44] $\times 10^{-1}$ \\
\hline
\end{tabular}
\end{table}

\begin{table}
\caption{The same as Table XX but for statistical significance at $5 \sigma$.}
\begin{tabular}{|c|c|c|c|c|c|}\hline
Couplings (TeV$^{-4}$) & 100 fb$^{-1}$ & 300 fb$^{-1}$ & 500 fb$^{-1}$ & 1000 fb$^{-1}$ \\
\hline \hline
$f_{M0}/\Lambda^{4}$  & [-1.27;1.27] & [-1.18;1.17]  & [-1.14;1.14]  & [-1.11;1.11]  \\
\hline
$f_{M1}/\Lambda^{4}$  & [-4.93;4.92] & [-4.55;4.54]  & [-4.42;4.41] & [-4.30;4.29]   \\
\hline
$f_{M7}/\Lambda^{4}$  & [-9.22;9.25]  & [-8.51;8.53]  & [-8.28;8.30]  & [-8.04;8.06]    \\
\hline
$f_{S0}/\Lambda^{4}$  &  [-3.06;3.03] $\times 10^{1}$ & [-2.83;2.79] $\times 10^{1}$
 & [-2.75;2.72] $\times 10^{1}$ & [-2.67;2.64]$\times 10^{1}$  \\
\hline
$f_{S1}/\Lambda^{4}$  & [-1.69;1.68]$\times 10^{1}$  & [-1.56;1.55]$\times 10^{1}$ & [-1.52;1.51]$\times 10^{1}$ & [-1.48;1.46]$\times 10^{1}$  \\
\hline
$f_{T0}/\Lambda^{4}$  &  [-1.98;2.01] $\times 10^{-1}$ & [-1.83;1.86] $\times 10^{-1}$ & [-1.78;1.80] $\times 10^{-1}$ & [-1.73;1.75] $\times 10^{-1}$ \\
\hline
$f_{T1}/\Lambda^{4}$  & [-3.06;2.98] $\times 10^{-1}$  & [-2.82;2.75] $\times 10^{-1}$ & [-2.75;2.68] $\times 10^{-1}$ & [-2.67;2.60] $\times 10^{-1}$  \\
\hline
$f_{T2}/\Lambda^{4}$  & [-5.28;5.25] $\times 10^{-1}$ & [-4.88;4.85] $\times 10^{-1}$ & [-4.74;4.71] $\times 10^{-1}$ & [-4.61;4.58] $\times 10^{-1}$ \\
\hline
\end{tabular}
\end{table}

\begin{table}
\caption{For 10 TeV FCC-he with integrated luminosities of 100, 300, 500 and 1000 fb$^{-1}$ through $W^{+}W^{-}$ production semileptonic decay channel, limits on the anomalous $\frac{f_{M0}}{\Lambda^{4}}$, $\frac{f_{M1}}{\Lambda^{4}}$, $\frac{f_{M7}}{\Lambda^{4}}$, $\frac{f_{S0}}{\Lambda^{4}}$, $\frac{f_{S1}}{\Lambda^{4}}$, $\frac{f_{T0}}{\Lambda^{4}}$, $\frac{f_{T1}}{\Lambda^{4}}$ and $\frac{f_{T2}}{\Lambda^{4}}$ couplings by using statistical significance at $3 \sigma$. Here, the obtained anomalous quartic couplings are in units of TeV$^{-4}$.}
\begin{tabular}{|c|c|c|c|c|c|}\hline
Couplings  & 100 fb$^{-1}$ & 300 fb$^{-1}$ & 500 fb$^{-1}$ & 1000 fb$^{-1}$ \\
\hline \hline
$f_{M0}/\Lambda^{4}$  & [-2.82;2.64] & [-2.44;2.25] & [-2.31;2.12] & [-2.16;1.98] \\
\hline
$f_{M1}/\Lambda^{4}$  & [-1.07;1.05] $\times 10^{1}$ & [-9.21;8.98]  & [-8.69;8.46] & [-8.14;7.91]   \\
\hline
$f_{M7}/\Lambda^{4}$  & [-2.44;2.46]  $\times 10^{1}$  & [-2.09;2.11] $\times 10^{1}$  & [-1.97;1.99] $\times 10^{1}$  & [-1.85;1.87]$\times 10^{1}$    \\
\hline
$f_{S0}/\Lambda^{4}$  &  [-4.47;4.41] $\times 10^{1}$ & [-3.84;3.78] $\times 10^{1}$ & [-3.63;3.56] $\times 10^{1}$ & [-3.39;3.33] $\times 10^{1}$ \\
\hline
$f_{S1}/\Lambda^{4}$  & [-1.88;1.87] $\times 10^{1}$  & [-1.62;1.61]$\times 10^{1}$  & [-1.52;1.52]$\times 10^{1}$  & [-1.43;1.42]$\times 10^{1}$   \\
\hline
$f_{T0}/\Lambda^{4}$  &  [-2.77;2.78] $\times 10^{-1}$ & [-2.38;2.38] $\times 10^{-1}$ & [-2.25;2.25] $\times 10^{-1}$ & [-2.10;2.11] $\times 10^{-1}$ \\
\hline
$f_{T1}/\Lambda^{4}$  & [-4.62;4.58] $\times 10^{-1}$  & [-3.97;3.93] $\times 10^{-1}$ & [-3.75;3.70] $\times 10^{-1}$ & [-3.51;3.46] $\times 10^{-1}$  \\
\hline
$f_{T2}/\Lambda^{4}$  & [-7.82;7.75] $\times 10^{-1}$ & [-6.72;6.65] $\times 10^{-1}$ & [-6.34;6.27] $\times 10^{-1}$ & [-5.94;5.87] $\times 10^{-1}$ \\
\hline
\end{tabular}
\end{table}

\begin{table}
\caption{The same as Table XXII but for statistical significance at $5 \sigma$.}
\begin{tabular}{|c|c|c|c|c|c|}\hline
Couplings (TeV$^{-4}$) & 100 fb$^{-1}$ & 300 fb$^{-1}$ & 500 fb$^{-1}$ & 1000 fb$^{-1}$ \\
\hline \hline
$f_{M0}/\Lambda^{4}$  & [-3.34;3.15] & [-2.79;2.61]  & [-2.60;2.41] & [-2.39;2.20]  \\
\hline
$f_{M1}/\Lambda^{4}$  & [-1.27;1.25]$\times 10^{1}$ & [-1.06;1.04] $\times 10^{1}$ & [-9.83;9.60] & [-9.01;8.78]   \\
\hline
$f_{M7}/\Lambda^{4}$  & [-2.90;2.92]  $\times 10^{1}$  & [-2.41;2.43]$\times 10^{1}$  & [-2.24;2.26]$\times 10^{1}$  & [-2.05;2.07]$\times 10^{1}$    \\
\hline
$f_{S0}/\Lambda^{4}$  &  [-5.31;5.25] $\times 10^{1}$ & [-4.42;4.35] $\times 10^{1}$  & [-4.10;4.04] $\times 10^{1}$ & [-3.76;3.70]$\times 10^{1}$  \\
\hline
$f_{S1}/\Lambda^{4}$  & [-2.24;2.23] $\times 10^{1}$ & [-1.86;1.85]$\times 10^{1}$ & [-1.73;1.72]$\times 10^{1}$ & [-1.58;1.57]$\times 10^{1}$  \\
\hline
$f_{T0}/\Lambda^{4}$  &  [-3.30;3.30] $\times 10^{-1}$ & [-2.74;2.74] $\times 10^{-1}$ & [-2.54;2.55] $\times 10^{-1}$ & [-2.33;2.33] $\times 10^{-1}$ \\
\hline
$f_{T1}/\Lambda^{4}$  & [-5.49;5.45] $\times 10^{-1}$  & [-4.57;4.52] $\times 10^{-1}$ & [-4.24;4.20] $\times 10^{-1}$ & [-3.89;3.84] $\times 10^{-1}$  \\
\hline
$f_{T2}/\Lambda^{4}$  & [-9.30;9.22] $\times 10^{-1}$ & [-7.73;7.66] $\times 10^{-1}$ & [-7.18;7.11] $\times 10^{-1}$ & [-6.58;6.50] $\times 10^{-1}$ \\
\hline
\end{tabular}
\end{table}

\appendix
\section{Fit Functions for the total cross sections of the process $ep \rightarrow \nu_{e}W^{+}W^{-} j$}

Numerical calculations of the total cross sections for $\frac{f_{M1}}{\Lambda^{4}}$ ,  $\frac{f_{S1}}{\Lambda^{4}}$ and $\frac{f_{T1}}{\Lambda^{4}}$ at $\sqrt{s}=1.30$, $1.98$, $7.07$ and $10$ TeV are given in Tables XXIV-XXV. Here, $\sigma_{0}$ is the SM cross section of the process $ep \rightarrow \nu_{e}W^{+}W^{-} j$ where $W^{+}$ and $W^{-}$ decays pure leptonic and semi-leptonic. On the other hand $\sigma_{1}$ and $\sigma_{2}$ are interference and new physics term, respectively.

\begin{table}
\caption{$\sigma_{0}$, $\sigma_{1}$ and $\sigma_{2}$ values of $\frac{f_{M1}}{\Lambda^{4}}$ , $\frac{f_{S1}}{\Lambda^{4}}$, $\frac{f_{T1}}{\Lambda^{4}}$ for leptonic final state at $\sqrt{s}=1.30$, $1.98$, $7.07$ and $10$ TeV.}
\begin{center}
\begin{tabular}{|c|c|c|c|}
\hline
Couplings & $\sigma_{0}$ (pb) & $\sigma_{1}$ (pb) & $\sigma_{2}$ (pb)\\
\hline
\hline
$\sqrt{s}=1.30$ TeV \\
\hline
$\frac{f_{M1}}{\Lambda^{4}}$ & 1.07 $\times 10^{-4}$ & 3.78 $\times 10^{3}$ & 1.41 $\times 10^{14}$  \\
\hline
$\frac{f_{S1}}{\Lambda^{4}}$ & 1.07 $\times 10^{-4}$ & 1.23 $\times 10^{3}$ & 1.15 $\times 10^{14}$ \\
\hline
$\frac{f_{T1}}{\Lambda^{4}}$ & 1.07 $\times 10^{-4}$ & 1.45 $\times 10^{5}$ & 2.33 $\times 10^{16}$\\
\hline
\hline
$\sqrt{s}=1.98$ TeV  \\
\hline
$\frac{f_{M1}}{\Lambda^{4}}$ & 4.60 $\times 10^{-4}$ & 1.16 $\times 10^{3}$ & 4.98 $\times 10^{15}$  \\
\hline
$\frac{f_{S1}}{\Lambda^{4}}$ & 4.60 $\times 10^{-4}$ & 2.11 $\times 10^{3}$ & 2.96 $\times 10^{15}$  \\
\hline
$\frac{f_{T1}}{\Lambda^{4}}$ & 4.60 $\times 10^{-4}$ & 1.38 $\times 10^{6}$ & 1.00 $\times 10^{18}$\\
\hline
\hline
$\sqrt{s}=7.07$ TeV  \\
\hline
$\frac{f_{M1}}{\Lambda^{4}}$ & 3.60 $\times 10^{-3}$ & 2.92 $\times 10^{5}$ & 9.96 $\times 10^{18}$  \\
\hline
$\frac{f_{S1}}{\Lambda^{4}}$ & 3.60 $\times 10^{-3}$ & 7.82 $\times 10^{5}$ & 1.20 $\times 10^{18}$  \\
\hline
$\frac{f_{T1}}{\Lambda^{4}}$ & 3.60 $\times 10^{-3}$ & 1.99 $\times 10^{7}$ & 2.56 $\times 10^{21}$\\
\hline
\hline
$\sqrt{s}=10.0$ TeV  \\
\hline
$\frac{f_{M1}}{\Lambda^{4}}$ & 7.35 $\times 10^{-3}$ & 2.02 $\times 10^{6}$ & 1.89 $\times 10^{20}$  \\
\hline
$\frac{f_{S1}}{\Lambda^{4}}$ & 7.35 $\times 10^{-3}$ & 2.31 $\times 10^{6}$ & 1.61 $\times 10^{19}$  \\
\hline
$\frac{f_{T1}}{\Lambda^{4}}$ & 7.35 $\times 10^{-3}$ & 3.59 $\times 10^{8}$ & 5.02 $\times 10^{22}$\\
\hline
\end{tabular}
\end{center}
\end{table}

\begin{table}
\caption{$\sigma_{0}$, $\sigma_{1}$ and $\sigma_{2}$ values of $\frac{f_{M1}}{\Lambda^{4}}$ , $\frac{f_{S1}}{\Lambda^{4}}$, $\frac{f_{T1}}{\Lambda^{4}}$ for semileptonic final state at $\sqrt{s}=1.30$, $1.98$, $7.07$ and $10$ TeV.}
\begin{center}
\begin{tabular}{|c|c|c|c|}
\hline
Couplings & $\sigma_{0}$ (pb) & $\sigma_{1}$ (pb) & $\sigma_{2}$ (pb)\\
\hline
\hline
$\sqrt{s}=1.30$ TeV \\
\hline
$\frac{f_{M1}}{\Lambda^{4}}$ & 2.39 $\times 10^{-4}$ & 1.48 $\times 10^{4}$ & 5.14 $\times 10^{14}$  \\
\hline
$\frac{f_{S1}}{\Lambda^{4}}$ & 2.39 $\times 10^{-4}$ & 5.52 $\times 10^{3}$ & 5.51 $\times 10^{14}$ \\
\hline
$\frac{f_{T1}}{\Lambda^{4}}$ & 2.39 $\times 10^{-4}$ & 2.71 $\times 10^{5}$ & 8.12 $\times 10^{16}$\\
\hline
\hline
$\sqrt{s}=1.98$ TeV  \\
\hline
$\frac{f_{M1}}{\Lambda^{4}}$ & 1.01 $\times 10^{-3}$ & 1.86 $\times 10^{5}$ & 1.13 $\times 10^{16}$  \\
\hline
$\frac{f_{S1}}{\Lambda^{4}}$ & 1.01 $\times 10^{-3}$ & 3.10 $\times 10^{4}$ & 9.52 $\times 10^{15}$  \\
\hline
$\frac{f_{T1}}{\Lambda^{4}}$ & 1.01 $\times 10^{-3}$ & 2.25 $\times 10^{5}$ & 2.30 $\times 10^{18}$\\
\hline
\hline
$\sqrt{s}=7.07$ TeV  \\
\hline
$\frac{f_{M1}}{\Lambda^{4}}$ & 8.41 $\times 10^{-3}$ & 1.99 $\times 10^{7}$ & 3.59 $\times 10^{18}$  \\
\hline
$\frac{f_{S1}}{\Lambda^{4}}$ & 8.41 $\times 10^{-3}$ & 4.53 $\times 10^{6}$ & 1.33 $\times 10^{18}$  \\
\hline
$\frac{f_{T1}}{\Lambda^{4}}$ & 8.41 $\times 10^{-3}$ & 9.77 $\times 10^{6}$ & 1.56 $\times 10^{21}$\\
\hline
\hline
$\sqrt{s}=10$ TeV  \\
\hline
$\frac{f_{M1}}{\Lambda^{4}}$ & 1.62 $\times 10^{-2}$ & 2.66 $\times 10^{7}$ & 1.97 $\times 10^{19}$  \\
\hline
$\frac{f_{S1}}{\Lambda^{4}}$ & 1.62 $\times 10^{-2}$ & 2.63 $\times 10^{6}$ & 5.93 $\times 10^{18}$  \\
\hline
$\frac{f_{T1}}{\Lambda^{4}}$ & 1.62 $\times 10^{-2}$ & 2.38 $\times 10^{8}$ & 9.54 $\times 10^{21}$\\
\hline
\end{tabular}
\end{center}
\end{table}


\begin{thebibliography}{99}

\bibitem{higgs1} S. Chatrchyan $et \: al.$, CMS collaboration, Phys. Lett. B 716, 30 (2012).
\bibitem{higgs2} G. Aad $et \: al.$, ATLAS collaboration, Phys. Lett. B 716, 1 (2012).
\bibitem{bel1} G. Belanger and F. Boudjema, Phys. Lett. B 288, 201 (1992).
\bibitem{bel2} G. Belanger, F. Boudjema, Y. Kurihara, D. Perret-Gallix and A. Semenov, Eur. Phys. J. C
13, 283 (2000)
\bibitem{baa} M. Baak $et \: al.$, The Snowmass EW WG report, arXiv:1310.6708 (2013).
\bibitem{x1} M. Koksal, Mod. Phys. Lett. A 29, 1450184 (2014).
\bibitem{x2} M. Koksal, Eur. Phys. J. Plus 130, 75 (2015).
\bibitem{x3} M. Koksal and A. Senol, Int. J. Mod.Phys. A 30, 1550107 (2015).
\bibitem{x4} A. Senol and M. Koksal, JHEP 1503, 139 (2015).
\bibitem{x5} A. Senol and M. Koksal, Phys. Lett. B 742, 143-148 (2015).
\bibitem{x6} A. Senol, M. Koksal and S. C. Inan, Adv.High Energy Phys. 2017, 6970587 (2017).
\bibitem{x7} M. Koksal, A. Senol and V. Ari, Adv.High Energy Phys. 2016, 8672391 (2016).
\bibitem{x8} A. Gutierrez-Rodriguez, C.G. Honorato, J. Montano, M.A. Perez, Phys. Rev. D 89, 034003
(2014).
\bibitem{x9} S. Fichet, G. von Gersdorff, O. Kepka, B. Lenzi, C. Royon, M. Saimper, Phys.Rev. D 89, 114004 (2014).
\bibitem{x91} S. Fichet, G. von Gersdorff, O. Kepka, B. Lenzi, C. Royon, M. Saimper, JHEP 1502, 165 (2015).
\bibitem{x92} C. Baldenegro, S. Fichet, G. von Gersdorff, C. Royon, JHEP 1706, 142 (2017).
\bibitem{x10} K. Ye, D. Yang and Q. Li, Phys. Rev. D 88, 015023 (2013).
\bibitem{x11} O. J. P. Eboli, M. C. Gonzalez-Garcia and S. F. Novaes, Nucl. Phys. B 411, 381 (1994).
\bibitem{x12} O. J. P. Eboli, M. B. Magro, P. G. Mercadante and S. F. Novaes, Phys. Rev. D 52, 15 (1995).
\bibitem{x13} O. J. P. Eboli and M. C. Gonzalez-Garcia, Phys.Rev. D 93, no.9, 093013 (2016).
\bibitem{x14} O. J. P. Eboli and J. K. Mizukoshi, Phys.Rev. D 64, 075011 (2001).
\bibitem{x141} O. J. P. Eboli, M. C. Gonzalez-Garcia and S. M. Lietti, Phys.Rev. D 69, 095005 (2004).
\bibitem{x15} M. Beyer et al., Eur. Phys. J. C 48, 353 (2006).
\bibitem{x16} T. Pierzchala and K. Piotrzkowski, Nucl. Phys. Proc. Suppl. 179180, 257 (2008).
\bibitem{x17} S. Atag, I. Sahin, Phys.Rev. D 70, 053014 (2004).
\bibitem{x18} S. Atag, I. Sahin, Phys.Rev. D 75, 073003 (2007).
\bibitem{x19} I. Sahin and B. Sahin, Phys.Rev. D 86, 115001 (2012).
\bibitem{x21} G. Perez, M. Sekulla and D. Zeppenfeld,  Eur.Phys. J. C 78  no.9, 759 (2018).
\bibitem{x22} A.S. Kurova and E.Yu. Soldatov, Phys. Atom. Nucl. 80 no.4, 725 (2017).
\bibitem{x33} J. Kalinowski $et \: al.$, Eur. Phys. J. C 78, 403 (2018).
\bibitem{yy} G. Perez, M. Sekulla, D. Zeppenfeld, Eur. Phys. J. C 78, 759 (2018).
\bibitem{zz} A. I. Ahmadov, arXiv:1806.03460.
\bibitem{pp} A. Senol, Int. J. Mod.Phys. A 29, 1450148 (2014).
\bibitem{sir1} A. M. Sirunyan et al, PHYSICAL REVIEW D 100, 012004 (2019).
\bibitem{sir2} A. M Sirunyan et al., CMS Collaboration, Phys.Lett.B 798 (2019) 134985.

\bibitem{sir3} A. M Sirunyan et al., CMS Collaboration, Phys.Lett.B 795 (2019) 281-307.

\bibitem{sir4} A. M Sirunyan et al., CMS Collaboration, Phys. Lett. B 774 (2017) 682-705

\bibitem{ma1} M. Aaboud et al., ATLAS Collaboration, Eur. Phys. J. C 77 (2017) 9, 646.

\bibitem{ma2} M. Aaboud et al., ATLAS Collaboration, JHEP 07 (2017) 107.

\bibitem{sir5} A. M Sirunyan et al., CMS Collaboration, JHEP 10 (2017) 072.

\bibitem{ma3} M. Aaboud et al., ATLAS Collaboration, Phys.Rev.D 96 (2017) 1, 012007.

\bibitem{11} O. J. P. Eboli, M. C. Gonzalez-Garcia, and J. K. Mizukoshi,Phys. Rev. D 74, 073005 (2006).
\bibitem{12} M. Fabbrichesi, M. Pinamonti, A. Tonero and A. Urbano, Phys. Rev. D 93, 015004 (2016).

\bibitem{atlas} M. Aaboud et al., ATLAS Collaboration, Eur. Phys. J. C 77, 141  (2017).
\bibitem{ebo} O. J. P. Eboli, M. C. Gonzalez-Garcia and J. K. Mizukoshi, Phys. Rev. D 74, 073005 (2006).
\bibitem{deg} C. Degrande et al., arXiv:1309.7452. (2013).
\bibitem{yu} Y. Wen, H. Qu, D. Yang, Q. s. Yan, Q. Li and Y. Mao, JHEP 1503, 025 (2015).
\bibitem{lhec}  H.Y. Bi, R. Y. Zhang, X. G. Wu, W. G. Ma, X. Z. Li and S. Owusu, Phys. Rev. D 95, 074020 (2017).
\bibitem{lhec1} J. L. A. Fernandez, et al., [LHeC Study Group], J. Phys. G39, 075001 (2012). 
\bibitem{lhec2} J. L. A. Fernandez, et al., [LHeC Study Group], arXiv:1211.5102. 
\bibitem{lhec3} O. Bruning and M. Klein, Mod. Phys. Lett. A28, 1330011 (2013). 
\bibitem{lhec4} A Large Hadron electron Collider at CERN, web page with recent papers, talks and workshop documentation: http://cern.ch/lhec.
\bibitem{lhec5} P. Agostini, et al., [LHeC Collaboration and FCC-he Study Group], arXiv:2007.14491 [hepex]. 
\bibitem{fcc} Y. C. Acar, A. N. Akay, S. Beser, H. Karadeniz, U. Kaya, B. B. Oner, S. Sultansoy, Nuclear
Inst. and Methods in Physics Research, A 871, 47-53 (2017).

\bibitem{fcc1} A. Abada et al. (FCC Collaboration), Eur. Phys. J. C 79, 474 (2019).

\bibitem{fcc2} A. Abada et al. (FCC Collaboration), Eur. Phys. J. Spec. Top. 228, 755 (2019).

\bibitem{rul} A. Alloul, N. D. Christensen, C. Degrande, C. Duhr, and B. Fuks, Computer Physics Communications,
185, 2250 (2014).
\bibitem{mad} J. Alwall, M. Herquet, F. Maltoni, O. Mattelaer, and T. Stelzer, Journal of High Energy
Physics, vol. 06, 128 (2011).
\bibitem{ufo} C. Degrande, C. Duhr, B. Fuks, D. Grellscheid, O. Mattelaer and T. Reiter, Comput. Phys. Commun.
183, 1201 (2012).
\bibitem{cteq} J. Pumplin, D. R. Stump, J. Huston, H. L. Lai, P. M. Nadolsky and W. K. Tung, JHEP 0207
012 (2002).

\bibitem{19-1} A. M Sirunyan et al., CMS Collaboration, CMS PAS SMP-19-008.

\bibitem{19-2} A. M Sirunyan et al., CMS Collaboration, CMS PAS SMP-18-007.

\bibitem{19-3} A. M Sirunyan et al., CMS Collaboration, Phys. Lett. B 809 (2020) 135710, 10.1016/j.physletb.2020.135710.

\bibitem{19-4} A. M Sirunyan et al., CMS Collaboration, Phys. Rev. D 100, 012004 (2019).

\bibitem{19-5} A. M Sirunyan et al., CMS Collaboration, CMS-SMP-20-001, CERN-EP-2020-127.

\bibitem{5-1} R. D. Ballet al.[NNPDF Collaboration], Nucl. Phys. B877, 290 (2013) [arXiv:1308.0598 [hep-ph]].

\bibitem{5-2} S. Carrazza [NNPDF Collaboration], PoS DIS2013, 279 (2013) [arXiv:1307.1131 [hep-ph]].

\bibitem{5-3} S. Carrazza [NNPDF Collaboration], arXiv:1305.4179 [hep-ph].

\bibitem{5-4} R. D. Ballet al.[NNPDF Collaboration], arXiv:1410.8849[hep-ph].

 \bibitem{9-1} T. Sjöstrand, P. Skands, JHEP 0403:053,2004,
 10.1088/1126-6708/2004/03/053.
 
 \bibitem{web} O.J.P. Eboli, M.C. Gonzalez-Garcia, http://feynrules.irmp.ucl.ac.be/wiki/AnomalousGaugeCoupling 

\end{thebibliography}
\end{document}